\documentclass[twocolumn,aps,pre,floatfix,superscriptaddress,longbibliography]{revtex4-1}
\pdfoutput=1 
\usepackage{amsmath,amssymb,eucal,graphicx,float,epstopdf,xparse}
\usepackage{epsfig,subfigure}
\usepackage[utf8]{inputenc}

\usepackage[colorlinks=true, urlcolor=blue, anchorcolor=blue, citecolor=blue,filecolor=blue,linkcolor=blue,menucolor=blue,pagecolor=blue]{hyperref}

\begin{document}

\author{R. R. Dyachenko}
\affiliation{Faculty of Computational Mathematics and Cybernetics, Lomonosov MSU, Moscow, Russia}
\affiliation{Institute of Numerical Mathematics RAS, Moscow, Russia}
\author{S. A. Matveev}
\affiliation{Faculty of Computational Mathematics and Cybernetics, Lomonosov MSU, Moscow, Russia}
\affiliation{Institute of Numerical Mathematics RAS, Moscow, Russia}
\author{P. L. Krapivsky}
\affiliation{Department of Physics, Boston University, Boston, Massachusetts 02215, USA}
\affiliation{Santa Fe Institute, Santa Fe, New Mexico 87501, USA}

\title{Finite Size Effects in  Addition and Chipping Processes}

\begin{abstract}
We investigate analytically and numerically a system of clusters evolving via collisions with clusters of minimal mass (monomers). Each collision either leads to the addition of the monomer to the cluster or the chipping of a monomer from the cluster, and emerging behaviors depend on which of the two processes is more probable. If addition prevails, monomers disappear in a time that scales as $\ln N$ with the total mass $N\gg 1$, and the system reaches a jammed state. When chipping prevails, the system remains in a quasi-stationary state for a time that scales exponentially with $N$, but eventually, a giant fluctuation leads to the disappearance of monomers. In the marginal case, monomers disappear in a time that scales linearly with $N$, and the final supercluster state is a peculiar jammed state, viz., it is not extensive. 
\end{abstract}

\maketitle

\section{Introduction}
\label{sec:intro}

Aggregation and fragmentation processes describing the merging and breaking  of clusters are widespread in nature \cite{Flory,Friedlander,MBE,KRB}. In aggregation, clusters can merge upon contact with the rate depending on the masses of the reactants (we ignore that chemical reactions often require catalysts). In fragmentation, clusters break up either by external driving or mutual collisions. Competition between aggregation and fragmentation often results in a steady state. 

In applications, clusters often represent a union of an integer number of elementary blocks known as monomers. Polymers composed of repeated subunits constitute a prime example hinting that a few types of monomers could be present, e.g., four in RNA  or DNA. We consider the simplest setting with one type of monomers.  Denote by $\mathbb{I}_k$ a cluster of mass $k$, i.e., composed of $k$ monomers. We thus tacitly assume that each cluster is fully described by its mass.

Addition is an aggregation process in which clusters grow by adding monomers. The simplest composite objects, dimers, arise via the reaction process $\mathbb{A}+\mathbb{A}\to \mathbb{I}_2$, where $\mathbb{A} = \mathbb{I}_1$ denotes a monomer. Trimers are formed by adding monomers to dimers, $\mathbb{A}+\mathbb{I}_2\to \mathbb{I}_3$,  and generally 
\begin{equation}
\label{scheme:add}
\mathbb{A}+\mathbb{I}_k\mathop{\longrightarrow}^{\mathrm{A}_k}  \mathbb{I}_{k+1}
\end{equation}
The addition process \eqref{scheme:add} provides a natural description of systems with mobile monomers and immobile composite objects (islands). One important application is to surface science where the monomers are adatoms hopping on the substrate \cite{MBE}. When two adatoms meet they form an {\em immobile} island, a dimer; similarly when an adatom meets an island $\mathbb{I}_k$, it attaches irreversibly forming an island $\mathbb{I}_{k+1}$ of mass $k+1$.  

Chipping is a binary fragmentation process in which one of the two fragments is the monomer. Aggregation and chipping processes exhibit intriguing behaviors; see \cite{chipping-KR,chipping-Satya98,chipping-Satya01,chipping-Satya01,chipping-Barma} for derivations, extensions, and applications. The generic duality between aggregation and fragmentation specializes in the duality between addition and chipping. Addition and chipping (AC) processes naturally occur if only monomers are mobile, and a monomer-cluster collision leads either to adding the monomer or chipping a monomer from the cluster. 

In AC processes, a monomer-island collision either results in addition \eqref{scheme:add} or leads to chipping:
\begin{equation}
\label{scheme:chip}
\mathbb{A} + \mathbb{I}_{k}\mathop{\longrightarrow}^{\mathrm{C}_k}  \mathbb{A} + \mathbb{A} + \mathbb{I}_{k-1}, \quad k\geq 3
\end{equation}
When $k=2$, only monomers remain after chipping: 
\begin{equation}
\label{scheme:chip2}
\mathbb{A} + \mathbb{I}_{2}\mathop{\longrightarrow}^{\mathrm{C}_2}  \mathbb{A} + \mathbb{A} + \mathbb{A}
\end{equation}
We thus consider collision-controlled chipping processes: A monomer can break off an island after a free monomer hits this island. (The AS processes with spontaneous chipping were considered, e.g., in Ref.~\cite{Blackman94}.)

The collection of addition rates $\mathrm{A}_k$ and chipping rates $\mathrm{C}_k$ fully specify the AC process. The AC process with mass-independent rates admits a natural reformulation asserting that every collision between a monomer and an island is productive: Addition occurs with probability $p$, and chipping occurs with probability $1-p$.  Thus
\begin{equation}
\label{scheme}
\mathbb{A}+\mathbb{I}_k\longrightarrow
\begin{cases}
\mathbb{I}_{k+1}                                             & \text{prob}  ~~p\\
\mathbb{A} + \mathbb{A} + \mathbb{I}_{k-1}  & \text{prob}  ~~1-p
\end{cases}
\end{equation}
for $k\geq 2$. A collision between two monomers is exceptional since chipping is impossible; addition $\mathbb{A} + \mathbb{A}\to \mathbb{I}_2$ still occurs with probability $p$.

If addition prevails, $p>1/2$, monomers quickly disappear and the evolution stops. The final monomer-free state is jammed. In the critical regime, $p=p_c=1/2$, addition and chipping processes almost balance each other, but the $\mathbb{A} + \mathbb{A}\to \mathbb{I}_2$ channel still leads to the disappearance of monomers. If chipping prevails, $p<1/2$, the system quickly falls into a universal (independent on the initial conditions) steady state. A jammed state arising when $p>1/2$ depends on the initial condition. The relaxation is exponential in time in the jammed and steady state regimes. In the critical regime, the decay is algebraic in time.  

The above results characterize an infinite system. An ultimate fate of a finite system is different, particularly when $p<1/2$. A steady state $p<1/2$ regime arising in an infinite system is not eternal, a giant fluctuation eventually leads to the extinction of monomers. Hence jamming is inevitable in finite AC processes. 


In Sec.~\ref{sec:AP}, we recapitulate the basic properties of the pure addition process ($p=1$). The analysis of the AC processes in the general case when $0 < p < 1$ relies on similar tools. Also, the results for the pure addition process shed light on the behaviors of the AC process in the jamming regime, $p>1/2$. 

In Sec.~\ref{sec:ACP}, we study the AC process in the infinite system. We employ a mean-field approach, i.e., we neglect correlations. Thus collisions occur with rates proportional to the product of the concentrations of reactants. Mathematically, the problem is described by an infinite set of coupled ordinary differential equations for cluster densities. 

In Sec.~\ref{sec:ACP-finite}, we consider finite systems. Jammed states are absorbing states, and any finite system gets jammed with probability one. The time to reach a jammed states scales according to
\begin{equation}
\label{lifetime}
T \propto
\begin{cases}
\ln N             &  p>\frac{1}{2}\\
N                  & p=\frac{1}{2}\\
e^{A(p) N}    & p<\frac{1}{2}\
\end{cases}
\end{equation}
with the total number $N$ of monomers.

\begin{figure}[h]
    \centering
\includegraphics[width=9cm]{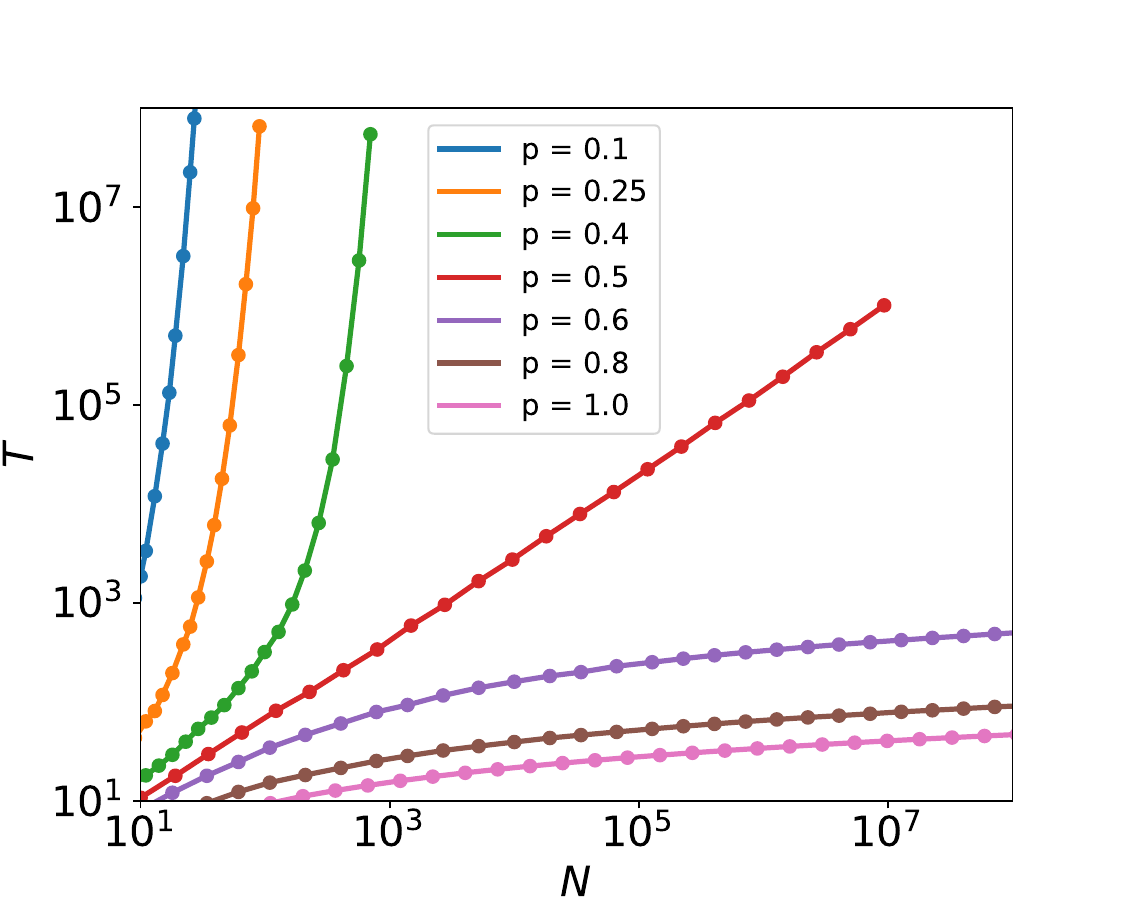}
\caption{The average lifetime $T$ versus $N$ for various values of $p$. Fits with Eq.~\eqref{lifetime} are present with dots. Theoretical predictions \eqref{lifetime} qualitatively agree with simulation results when $p\leq 1/2$. When $p > 1/2$, a non-linear logarithmic growth $T \propto (\ln N)^{\alpha(p)}$  with exponent satisfying $\alpha(p)>1$ and varying with $p$ better fits the data. For each $N$, we used $10^{3}$ Monte Carlo runs to estimate the average lifetime.}
\label{Fig:Mean-time}
\end{figure}

We use an efficient Monte Carlo algorithm \cite{Os:LRMC} to simulate the AC process in a finite system. As  an illustration, in Fig.~\ref{Fig:Mean-time}, we compare simulation results and theoretical predictions \eqref{lifetime}. We mostly rely on simulations to study the behavior in the quasi-stationary regime, $p<1/2$. The amplitude $A(p)$ in \eqref{lifetime} is unknown, but it vanishes when $p\uparrow 1/2$, and hence the direct Monte Carlo simulations allow us to reach a jammed state when $1/2-p$ is sufficiently small. The behavior in the jamming regime, $p>1/2$, essentially follows from the behavior of an infinite system, so it is analytically accessible. 

In the critical regime, the late stage is fluctuation-dominated, so it is not captured by the kinetic equations describing an infinite system. We probe fluctuations analytically in Sec.~\ref{sec:F} using the van Kampen expansion \cite{VanKampen,KRB}. The final state significantly varies from realization to realization.  For instance, the total number of clusters $\mathcal{C}$ is a non-self-averaging random variable exhibiting a non-extensive (sub-linear) scaling with $N$:
\begin{equation}
\label{clusters:crit}
\mathcal{C} \sim N^{4/5}  
\end{equation}
The typical cluster mass in the jammed state scales as 
\begin{equation}
\label{mass:crit}
k_\text{typ} \sim N^{1/5}  
\end{equation}

Similar supercluster jammed states composed of predominantly large clusters have been detected in addition and shattering processes \cite{Wendy21} in the critical regimes. 

The AC processes with proportional rates, $\mathrm{C}_k = \lambda \mathrm{A}_k$, behave similarly to the processes with mass-independent rates---the outcome depends on whether addition or chipping prevails. In Sec.~\ref{sec:prop} and Appendices \ref{ap: AC_linear}--\ref{ap: AC_alg}, we analyze the AC processes with proportional rates that vary algebraically, $\mathrm{A}_k=k^a$. For this class of models, the supercluster state is again rather peculiar, e.g.,  the total number of clusters always scales sub-linearly: 
\begin{equation}
\label{number:a}
\mathcal{C} \sim N^\frac{4-a}{5-a}
\end{equation}

The derivations of the scaling laws \eqref{clusters:crit}--\eqref{number:a} and other results about supercluster states are not rigorous as we rely on the van Kampen expansion beyond the range where it is formally exact. Thus one cannot determine the amplitudes, but the exponents in the scaling laws are believed to be exact. We discuss these caveats in Sec.~\ref{sec:disc}, and we also provide more details for most tractable versions, namely for the model with mass-independent rates (Sec.~\ref{sec:F}) and linear in mass rates (Appendix \ref{ap: AC_linear}).

\section{Addition Process}
\label{sec:AP}

When $p=1$, the AC process reduces to the simplest addition process. The governing equations read \cite{BK91,Blackman91,KRB}
\begin{subequations}
\begin{align}
&\frac{d c_1}{dt}=-c_1(c_1+c),\quad c=\sum_{j\geq 1} c_j
\label{1t}\\
&\frac{d c_k}{dt}=c_1(c_{k-1}-c_k), \quad k\geq 2
\label{2t}
\end{align}
\end{subequations}
Here $c_k$ is the density of clusters of mass $k$, so $k=1$ corresponds to mobile monomers, and $k\geq 2$ describe immobile islands. Using \eqref{1t}--\eqref{2t}, one can verify that the mass density $\sum_{j\geq 1} jc_j$ remains constant. In the following, we always set mass density to unity:
\begin{equation}
\label{mass}
\sum_{j\geq 1} jc_j=1
\end{equation}

Summing \eqref{1t} and all equations \eqref{2t} we obtain a rate equation for the total cluster density 
\begin{equation}
\label{Nt}
\frac{d c}{dt}=-c_1 c
\end{equation}
Introducing the auxiliary time
\begin{equation}
\label{time}
\tau=\int_0^t dt'\, c_1(t')
\end{equation}
we reduce Eqs.~(\ref{1t})--(\ref{2t}) to a set of linear equations
\begin{subequations}
\label{123}
\begin{align}
&\frac{d c_1}{d\tau}=-c_1-c
\label{1}\\
&\frac{d c_k}{d\tau}=c_{k-1}-c_k, \quad k\geq 2
\label{2}\\
&\frac{d c}{d\tau}=-c
\label{3}
\end{align}
\end{subequations}
For convenience, we also added \eqref{3} which is the reduced form of Eq.~\eqref{Nt}.  

Linear equations \eqref{1}--\eqref{3} are solvable for arbitrary initial conditions. In the following, we consider the most natural mono-disperse initial condition 
\begin{equation}
\label{ckt=0}
c_k(t=0)=\delta_{k,1}
\end{equation}
if not stated otherwise. 

Solving equation (\ref{3}) we find the cluster density and then from \eqref{1} we deduce the monomer density:
\begin{equation}
\label{Nc1:add}
c(\tau)=e^{-\tau}, \quad c_1(\tau)=(1-\tau)\,e^{-\tau}
\end{equation}
Using the monomer density we solve Eqs.~\eqref{2} recursively and find all island densities \cite{BK91}:
\begin{equation}
\label{ckt-sol}
c_k(\tau)=\left\{\frac{\tau^{k-1}}{(k-1)!}-\frac{\tau^k}{k!}\right\}e^{-\tau},\quad k\geq 1
\end{equation}
When $t\to\infty$ (this corresponds to $\tau\to 1$ for the mono-disperse initial conditions), the densities become
\begin{equation}
\label{ck-final}
C_k =\frac{k-1}{e\cdot k!}
\end{equation}
Hereinafter we use capital letters for final densities, so $C_k\equiv  c_k(t=\infty)$. The approach to the final state is exponential, e.g., the density of adatoms vanishes as 
\begin{equation}
\label{c1t}
c_1 \sim e^{-t/e}
\end{equation}
which follows from \eqref{1t} and $C\equiv c(t=\infty)=1/e$.  Combining the exact expression $c_1(\tau)=(1-\tau)\,e^{-\tau}$ and \eqref{time} one can extract a more precise asymptotic
\begin{equation}
\label{c1:asymp}
c_1 \simeq Ae^{-t/e}, \quad A=e^{-\gamma-1+\text{Ei}(-1)}=0.1658619\ldots
\end{equation}
where $\gamma=0.5772\ldots$ is the Euler constant and 
\begin{equation}
\text{Ei}(z)=\int_{-\infty}^z\frac{dx}{x}\,e^{x}
\end{equation}
is the exponential integral function.

\section{Addition and Chipping}
\label{sec:ACP}

For the AC process, we use again the time variable \eqref{time} to linearize the governing equations:
\begin{subequations}
\label{AC:eq}
\begin{align}
&\frac{d c_1}{d\tau}= -p(c_1 + c) + (1-p)(c_2-c_1+c)
\label{1AC}\\
&\frac{d c_k}{d\tau}=pc_{k-1}-c_k + (1-p) c_{k+1}, \quad k\geq 2
\label{2AC}\\
&\frac{d c}{d\tau}= -(1-p)c_1 + (1-2p)c
\label{3AC}
\end{align}
\end{subequations}
The rate equation \eqref{3AC} for the cluster density is obtained by summing \eqref{1AC} and all equations \eqref{2AC}; thus, it is not independent but convenient for future analysis. 

\begin{figure}
\centering
\includegraphics[width=7.77cm]{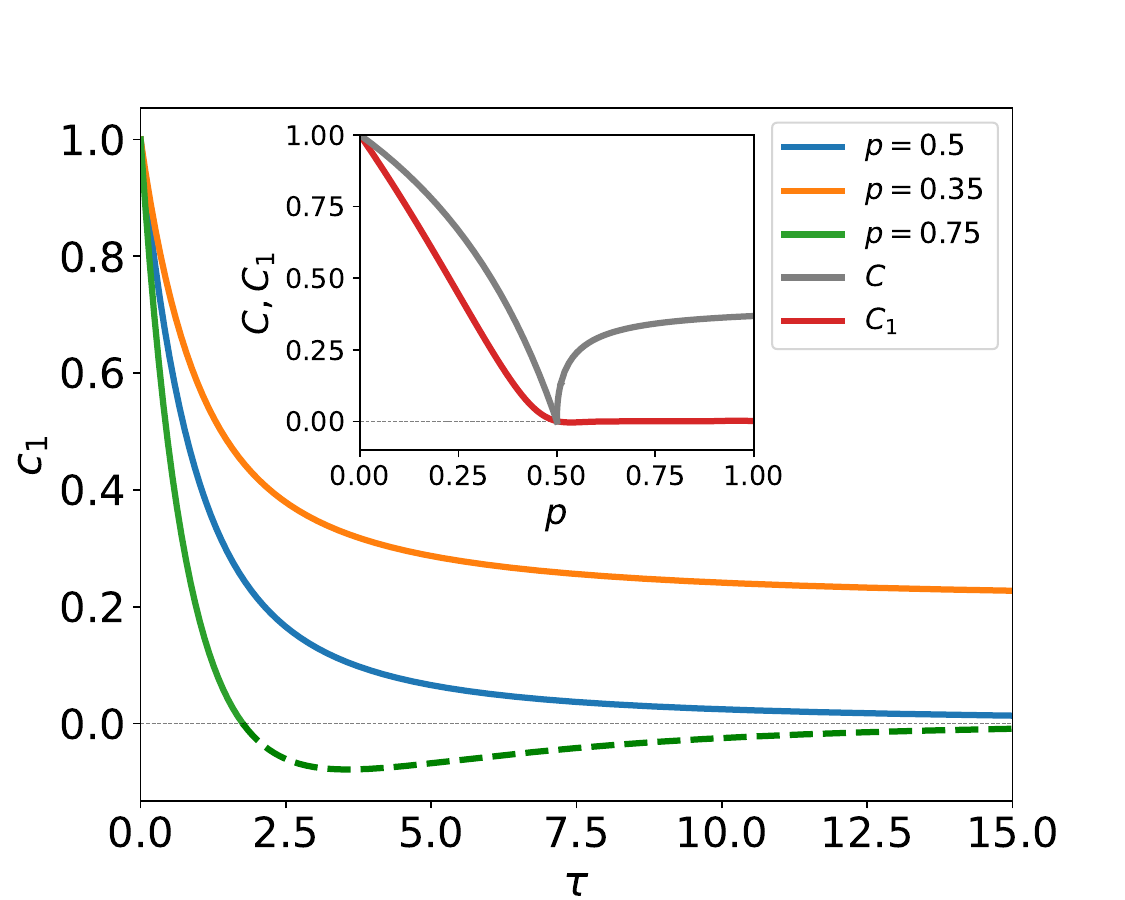}
\caption{The temporal decay of the monomer density  $c_1(\tau)$ for $p=7/20, ~p=1/2, ~p=3/4$  (top to bottom) exemplifying the evolution in the steady state, critical, and jamming regimes. In the critical regime, the monomer density is given by explicit formula \eqref{c1:RW-sol}. Generally the exact Laplace transform $\widehat{c}_1(s)$ is known, Eq.~\eqref{Lap:c1c}, and numerically inverting it we obtained $c_1(\tau)$ for $p=7/20$ and $p=3/4$. In the jamming regime, the monomer density vanishes at finite modified time $\tau_\text{max}(p)$ corresponding to $t=\infty$; e.g., $\tau_\text{max}(3/4)\approx 1.75703$. The inset shows that the final cluster density $C(p)$ and the final monomer density $C_1(p)$ undergo the continuous phase transition at $p_c=1/2$. The final monomer density, $C_1(p)$, is given by \eqref{final:mon}.  The final cluster density is known in the steady state regime, Eq.~\eqref{final:density}. In the jamming regime, the exact Laplace transform $\widehat{c}(s)$ is also known, Eq.~\eqref{Lap:c1c}. Instead of numerically inverting the Laplace transform and also finding $\tau_\text{max}(p)$ it is easier to determine  $C(p)$ using direct integration. }
\label{Fig:AC-monomers}
\end{figure}

Qualitatively different behaviors emerge depending on whether the probability $p$ is below, equal or above the critical value $p=p_c=1/2$ where addition and chipping are equiprobable (see Fig.~\ref{Fig:AC-monomers}). One can extract the time-dependent behavior of the densities $c_k(t)$ from an infinite set of linear differential equations \eqref{AC:eq} using the Laplace transform. This is laborious and the inversion of the exact Laplace transforms is usually impossible in terms of standard special functions.

The steady state emerges when $p<1/2$. In the steady state, Eqs.~\eqref{2AC} reduce to $pC_{k-1}-C_k + (1-p) C_{k+1}=0$. This recurrence admits an exponential solution  
\begin{equation}
\label{ckc:final}
C_k(p) = (1-2p)^2   \frac{p^{k-1}}{(1-p)^{k+1}}\,, \quad  C(p) = \frac{1-2p}{1-p}
\end{equation}

Thus the final monomer density in the entire $0<p\leq 1$ range reads 
\begin{equation}
\label{final:mon}
C_1(p) = 
\begin{cases}
0                                                            & p \geq  \frac{1}{2}\\
\left(\frac{1-2p}{1-p}\right)^2                  & p<\frac{1}{2}
\end{cases}
\end{equation}
The final monomer density undergoes a continuous phase transition as a function of $p$, see the inset in Fig.~\ref{Fig:AC-monomers}. 

The final cluster density also undergoes a continuous phase transition: 
\begin{equation}
\label{final:density}
C(p) = 
\begin{cases}
C_\infty(p)                               &p>\frac{1}{2}\\
0                                              &p=\frac{1}{2}\\
\frac{1-2p}{1-p}                        &p<\frac{1}{2}
\end{cases}
\end{equation}
The final cluster density $C_\infty(p)$ in the $p>1/2$ jamming regime is discussed below. Here we just mention that in the proximity of the critical point
\begin{equation}
\label{final:around}
C(p) \simeq 
\begin{cases}
2\sqrt{\frac{2p-1}{\pi}}         &0< p - \frac{1}{2} \ll 1\\
2(1-2p)                               &0< \frac{1}{2}-p \ll 1
\end{cases}
\end{equation}

\subsection{Critical regime:  $p = p_c=\frac{1}{2}$}
\label{subsec:critical}

In the critical regime, the final densities vanish. The time-dependent behavior is interesting and tractable. Equations \eqref{1AC}--\eqref{2AC} become
\begin{subequations}
\begin{align}
&2\frac{d c_1}{d\tau}= c_2 - 2c_1 
\label{1:crit}\\
&2\frac{d c_k}{d\tau}=c_{k-1}-2c_k + c_{k+1}, \quad k\geq 2
\label{2:crit}
\end{align}
\end{subequations}
It is convenient to consider 
\begin{equation}
\label{ck:RW}
2\frac{d c_k}{d\tau}=c_{k-1}-2c_k + c_{k+1}
\end{equation}
for all integer $k$. Setting the initial condition
\begin{equation}
\label{IC:dipole}
c_k(0)=\delta_{k,1}-\delta_{k,-1}
\end{equation}
we see that the solution of \eqref{ck:RW}--\eqref{IC:dipole} is an odd function of mass: $c_k(\tau)=-c_{-k}(\tau)$. Thus $c_0(\tau)\equiv 0$ and hence Eq.~\eqref{ck:RW} at $k=1$ turns into \eqref{1:crit}. Therefore the solution of \eqref{1:crit}--\eqref{2:crit} subject to the mono-disperse initial condition \eqref{ckt=0} coincides with the solution of the initial-value problem \eqref{ck:RW}--\eqref{IC:dipole}. 

To solve \eqref{ck:RW}--\eqref{IC:dipole} we observe that Eq.~\eqref{ck:RW} describes the probability distribution of a nearest-neighbor symmetric random walk on the one-dimensional lattice. For such random walk starting at the origin, the probability distribution is $e^{-\tau}I_k(\tau)$ where $I_k$ is the modified Bessel function of order $k$  (see, e.g., \cite{KRB}). For the ``dipole'' initial condition \eqref{IC:dipole} we therefore arrive at 
\begin{equation}
\label{ck:RW-sol}
c_k(\tau) = e^{-\tau}\left[I_{k-1}(\tau)-I_{k+1}(\tau)\right]
\end{equation}
Using identity $I_{\nu-1}(x)-I_{\nu+1}(x)=\frac{2\nu}{x}I_\nu(x)$ one can re-write \eqref{ck:RW-sol} through a single Bessel function
\begin{equation}
\label{ck:RW-sol-2}
c_k(\tau) = \frac{2k}{\tau}\,e^{-\tau}I_k(\tau)
\end{equation}
The monomer and cluster densities read
\begin{subequations}
\begin{align}
\label{c1:RW-sol}
& c_1(\tau) = \frac{2}{\tau}\,e^{-\tau} I_1(\tau)\\
\label{N:RW-sol}
& c(\tau) = e^{-\tau}\left[I_0(\tau) + I_{1}(\tau)\right]
\end{align}
\end{subequations}

Using the asymptotic $e^{-\tau}I_k(\tau)\simeq (2\pi \tau)^{-1/2}$ valid when $k=O(1)$ and $\tau\to \infty$, we deduce 
\begin{equation}
\label{c1N}
c_1(\tau)\simeq \sqrt{\frac{2}{\pi \tau^3}}, \quad  c(\tau)\simeq \sqrt{\frac{2}{\pi \tau}}
\end{equation}
from \eqref{c1:RW-sol}--\eqref{N:RW-sol}.  Therefore 
\begin{equation}
\label{t-tau}
t = \int_0^\tau \frac{d\tau' }{c_1(\tau')}\simeq \frac{1}{5}\,\sqrt{2\pi \tau^5}
\end{equation}
when $\tau\to \infty$. Re-writing \eqref{c1N} through the physical time we arrive at the large time behavior
\begin{subequations}
\begin{align}
\label{c1c:asymp}
c_1(t)\simeq \gamma\, t^{-3/5}, \quad  c(t)\simeq \nu\, t^{-1/5}
\end{align}
with
\begin{equation}
\label{gn:def}
\gamma = \left(\frac{16}{125 \pi}\right)^{1/5}, \qquad 
\nu = \left(\frac{8}{5 \pi^2}\right)^{1/5}
\end{equation}
\end{subequations}

We also note that the mass distribution \eqref{ck:RW-sol-2} acquires a simple scaling form
\begin{subequations}
\begin{equation}
\label{ck:RW-scaling}
c_k(\tau) \simeq \frac{2k}{\sqrt{2\pi \tau^3}}\,\exp\!\left[-\frac{k^2}{2\tau}\right]
\end{equation}
in the scaling limit
\begin{equation}
k\to\infty, \quad \tau \to\infty, \quad \frac{k}{\sqrt{\tau}} = \text{finite}
\end{equation}
\end{subequations}
In terms of the physical time, \eqref{ck:RW-scaling} becomes
\begin{subequations}
\begin{equation}
\label{ck:scaling}
c_k(t) \simeq \frac{\gamma k}{t^{3/5}}\,\exp\!\left[-\frac{\gamma}{2\nu}\,\frac{k^2}{t^{2/5}}\right]
\end{equation}
This mass distribution provides an asymptotically exact description in the scaling limit
\begin{equation}
k\to\infty, \quad t \to\infty, \quad \frac{k}{t^{2/5}} = \text{finite}
\end{equation}
\end{subequations}

\subsection{Laplace transform}
\label{subsec:LP}

The linearity of Eqs.~\eqref{AC:eq} suggests to apply the Laplace transform 
\begin{equation}
\label{Lap}
\widehat{c}_k(s) = \int_0^\infty d\tau\,e^{-s\tau} c_k(\tau)
\end{equation}
The Laplace transform recasts \eqref{2AC} into recurrence
\begin{equation}
\label{Lap:2AC}
s \widehat{c}_k = p\widehat{c}_{k-1}- \widehat{c}_k + (1-p)\widehat{c}_{k+1}, \quad k\geq 2
\end{equation}
which admits an exponential solution $\widehat{c}_k = A z^{k-1}$ with $z$ being a root of $(1-p) z^2 - (1+s)z + p=0$. An appropriate root giving the decaying (with mass) solution is
\begin{equation}
\label{z:root}
z = \frac{2p}{1+s+\sqrt{(1+s)^2-q^2}}\,, \quad q \equiv \sqrt{4p(1-p)}
\end{equation}
Applying the Laplace transform to the mass conservation relation $\sum_{k\geq 1} kc_k(\tau)=1$ one gets 
\begin{equation}
\label{Lap_mass}
\sum_{k\geq 1} k \widehat{c}_k(s) = \frac{1}{s}
\end{equation}
This sum rule fixes the amplitude $A=(1-z)^2/s$. Thus the Laplace transform is given by 
\begin{equation}
\label{Lap:sol}
\widehat{c}_k(s) = \frac{(1-z)^2}{s}\,z^{k-1}
\end{equation}
with $z$ determined by \eqref{z:root}. In particular,
\begin{equation}
\label{Lap:c1c}
\widehat{c}(s) = \frac{1-z}{s} \quad\text{and}\quad \widehat{c}_1(s) = \frac{(1-z)^2}{s}
\end{equation}
are the Laplace transforms of the total cluster density and the monomer density. 

The Laplace transforms \eqref{Lap:c1c} are neat, but inverting them in terms of special functions appears impossible.  One  can derive simple integral representations as we now show. Consider first the total cluster density. Using \eqref{z:root} and \eqref{Lap:c1c} we re-write $\widehat{c}(s)$ in the form
\begin{equation}
\label{Lap:c}
\widehat{c}(s) = \frac{1}{s}-\frac{2p}{s}\,\frac{1}{1+s+\sqrt{(1+s)^2-q^2}}
\end{equation}
To perform an inverse Laplace transform of the second term appearing on the right-hand side of \eqref{Lap:c} we rely on identity \cite{Bateman:Lap} 
\begin{subequations}
\begin{equation}
\label{Lap:Bessel}
\frac{1}{1+s+\sqrt{(1+s)^2-q^2}} \longrightarrow e^{-\tau}\,\frac{I_1(q\tau)}{q\tau}
\end{equation}
and a general property
\begin{equation}
\label{Lap:identity}
s^{-1} \widehat{f}(s)  \longrightarrow \int_0^\tau du\,f(u)
\end{equation}
\end{subequations}
With these ingredients, we derive an integral representation of the cluster density
\begin{subequations}
\begin{equation}
\label{c:IR}
c(\tau)=1-2p \int_0^\tau du\,e^{-u}\,\frac{I_1(q u)}{q u}
\end{equation}

To establish an integral representation of the monomer density, we use \eqref{3AC} and \eqref{c:IR} to give
\begin{equation}
\label{c1:IR}
c_1(\tau) = \frac{1-2p}{1-p}\,c(\tau)+\frac{2p}{1-p}\,e^{-\tau}\,\frac{I_1(q\tau)}{q\tau}
\end{equation}
\end{subequations}

These results apply for all $0 < p < 1$. For instance, $q=\sqrt{4p(1-p)}=1$ in the critical  $p = 1/2$ regime, so the integral in \eqref{c:IR} becomes
\begin{equation*}
\int_0^\tau du\,e^{-u}\,\frac{I_1(u)}{u} = 1 - e^{-\tau}\left[I_0(\tau) + I_{1}(\tau)\right]
\end{equation*}
and Eq.~\eqref{c:IR} reduces to \eqref{N:RW-sol}. For $p\ne 1/2$, however,  it does not seem possible to express the densities through known special functions. 

The inset in Fig.~\ref{Fig:AC-monomers} demonstrates that the final cluster density is continuous but loses smoothness at the critical point. The final cluster is known in the steady state regime, Eq.~\eqref{final:density}. To determine the final cluster in the jamming regime, it is in principle possible to use the exact Laplace transform $\widehat{c}(s)$, numerically invert it, and specialize to $\tau_\text{max}(p)$ implicitly determined by equation $c_1[\tau_\text{max}(p)]=0$. It is easier, however, to determine $C_\infty(p)=c[\tau_\text{max}(p)]$ using direct integration of Eqs.~\eqref{AC:eq}. 

Specifically, we apply the second-order Runge-Kutta time-integration scheme. We have verified that our numerical results are very precise by comparing with the analytical expression \eqref{c1:RW-sol} in the critical regime. In the jamming regime, numerical integration is even more precise and requires modest computing resources  as the densities are rapidly decaying with mass. The results for $C(p)$ in the jamming regime, $p>\frac{1}{2}$, shown in the inset in Fig.~\ref{Fig:AC-monomers} are obtained using numerical integration.

\subsection{Jamming regime:  $p > \frac{1}{2}$}

In the jamming regime, the maximal modified time $\tau_\text{max}(p)$ corresponding to infinite physical time is implicitly determined by $c_1(\tau_\text{max})=0$. The jammed cluster density is $C_\infty(p)=c[\tau_\text{max}(p)]$. It seems impossible to express $C_\infty(p)$ via known special functions. Here we deduce an explicit asymptotic behavior of $C_\infty(p)$ near the critical regime, $0<p - \frac{1}{2}\ll 1$. 

Writing $\epsilon=2p-1$, we get $q = \sqrt{4p(1-p)} = \sqrt{1-\epsilon^2}$.  Just above the critical point, $0<\epsilon\ll 1$, we expand \eqref{c1:IR} and find
\begin{equation}
\label{c1+}
c_1(\tau) = -2\epsilon c(\tau) + 2(1+2\epsilon)\, e^{-\tau}\,\frac{I_1(\tau)}{\tau} + O(\epsilon^2)
\end{equation}
In the leading order $c(\tau)=e^{-\tau}\left[I_0(\tau) + I_{1}(\tau)\right]$ which is just the cluster density in the critical regime, see \eqref{N:RW-sol}.  Plugging this into \eqref{c1+} and solving $c_1(\tau_\text{max})=0$ we find $\tau_\text{max}=(2\epsilon)^{-1}$ in  the leading order. Therefore
\begin{equation}
\label{C+}
C_\infty(p)=c(\tau_\text{max})\simeq \sqrt{\frac{2}{\pi \tau_\text{max}}}\simeq 2\sqrt{\frac{2p-1}{\pi}}
\end{equation}
as announced in \eqref{final:around}. 

To establish the large time decay of the monomer density we employ the same approach as in the derivation of \eqref{c1t} for the pure addition process and find an asymptotically exponential decay
\begin{equation}
\label{c1t-AC}
c_1 \sim e^{-Bt}, \qquad B = (2p-1)C-(1-p)C_2
\end{equation}
To determine the amplitude $B$ we need $C_2=c_2(\tau_\text{max})$ and $C=c(\tau_\text{max})$. Re-writing \eqref{1AC} as
\begin{equation}
\label{c2:IR}
(1-p)c_2 = \frac{d c_1}{d\tau} + c_1 + (2p-1)c
\end{equation}
and inserting \eqref{c:IR}--\eqref{c1:IR} into \eqref{c2:IR} we find $c_2(\tau)$ from which we extract $C_2=c_2(\tau_\text{max})$. Similarly from \eqref{c:IR} we get $C=c(\tau_\text{max})$. Thus we express $B$ via $\tau_\text{max}$.

\section{Finite Systems}
\label{sec:ACP-finite}

In a system with total mass $N$, a non-adsorbing state can dissolve into the disentangled state with $N$ monomers and no islands, and vice versa. Hence non-adsorbing states are mutually connected. Jammed states are absorbing, and each such state is connected to a few non-adsorbing states, so a finite system gets jammed with probability one. 

The road to jamming depends on whether the probability $p$ is below, equal, or above $p=p_c=1/2$. We now outline theoretical expectations of the behavior in these regimes. We focus on the average lifetime $T$ and briefly discuss the lifetime distribution. We also probe the behavior of the number of distinct island species and the number of islands in the final jammed state.

\subsection{Jamming regime:  $p > \frac{1}{2}$}
\label{subsec:FS-J}

The average total number of monomers is close to $M=Nc_1$, at least when $M$ is large. Neglecting fluctuations leads to the criterion $M(T)=Nc_1(T)\sim 1$ for estimating the average lifetime. Since the density decays exponentially, Eq.~\eqref{c1t-AC},  the average lifetime scales logarithmically with $N$ as stated in \eqref{lifetime}. One even expresses an amplitude through the decay rate $B$ in \eqref{c1t-AC}:
\begin{equation}
\label{T+:naive}
T \simeq B^{-1} \ln N
\end{equation}

In a finite system, the total number $\mathcal{C}_1$ of monomers is random. Fluctuations are relatively small in large systems, and they are traditionally investigated in the realm of the van Kampen expansion \cite{VanKampen}. In the present case one writes 
\begin{align}
\label{M:def}
\mathcal{C}_1(t) = N c_1(t) + \sqrt{N} \xi_1(t)
\end{align}
as the sum of the linear in $N$ deterministic contribution and proportional to $\sqrt{N}$ stochastic contribution, i.e., $\xi_1(t)=O(1)$ is a random variable. Van Kampen expansions have been used in the analyses of various reaction processes \cite{VanKampen,KRB,BK-van,McKane} including  aggregation and annihilation processes \cite{Lushnikov,Spouge85b,Sid86,Ernst87a,Ernst87b,Hilhorst04}. Assuming that the $\sqrt{N}$ scaling of fluctuations holds till the very end, we estimate the average lifetime $T$ from $Nc_1(T)\sim \sqrt{N}$. This gives 
\begin{equation}
\label{T+}
T \simeq (2B)^{-1} \ln N
\end{equation}
twice smaller than the naive estimate \eqref{T+:naive}. The above argument tacitly assumes that $\xi_1(t)$ remains of order one in the $t\to\infty$ limit. This expectation is erroneous in the critical regime (Sec.~\ref{subsec:FS-C}). In the jamming regime, however, the modified time is the natural variable. The evolution span is thus effectively finite, $\tau\leq  \tau_\text{max}$, and $\xi_1(\tau_\text{max})$ is expected to remain finite. 

A non-linear logarithmic growth law, $T \propto (\ln N)^{\alpha(p)}$, where the exponent is a decreasing function of $p$ satisfying $\alpha(p)>1$, provides a better fit to our simulation results than the linear logarithmic growth (see Fig.~\ref{Fig:Mean-time}). For instance, the best fit to simulation results for $p=0.6$ is $\alpha(0.6) =1.93$. For the pure addition process ($p=1$), the lifetime must grow as $\ln N$; simulations suggest $\alpha(1) =1.15$. 

It is feasible that $T \propto \ln N$ is the true asymptotic growth in the entire jamming regime,  $\frac{1}{2} < p < 1$. This asymptotic should be reached when $\ln N \gg 1$. Although the final state is logarithmically quickly reached the jamming regime, simulating astronomically large systems ($\ln N \gg 1$) is difficult. Finally, to appreciate why the fitting exponent  $\alpha(p)$ increases as $p\downarrow \frac{1}{2}$, recall that in the critical regime the lifetime scales as $T\sim N$, i.e., much faster than logarithmically. Thus when $0<p-\frac{1}{2}\ll 1$, it takes a long time before the system ``realizes" that it is not in the critical regime.

\subsection{Critical regime:  $p = \frac{1}{2}$}
\label{subsec:FS-C}

Recall that the monomer density is $c_1\sim t^{-3/5}$ in the critical regime in the infinite system. The naive criterion $M(T)=Nc_1(T)\sim 1$ gives an estimate $T \sim N^{5/3}$ for the average lifetime in the critical regime. We now argue that this naive estimate is erroneous not merely by a numerical factor as in the jamming regime [cf. \eqref{T+:naive} and \eqref{T+}]. The exponent is wrong, and instead of $T \sim N^{5/3}$ the lifetime scales linearly as announced in \eqref{lifetime}:
\begin{equation}
\label{T:crit}
T \sim N
\end{equation}

To establish \eqref{T:crit} we rely on the asymptotic behavior of the variance 
\begin{equation}
\label{xi-var}
\langle \xi_1^2 \rangle \sim c
\end{equation}
Arguments in favor of \eqref{xi-var} are presented in Section~\ref{sec:F}. The stochastic part of \eqref{M:def} scales as $\sqrt{N} \sqrt{\langle \xi_1^2 \rangle}\sim \sqrt{N c}$, while the deterministic part is $N c_1$. Equating these contributions, $Nc_1\sim \sqrt{N c}$, and using \eqref{c1c:asymp} we obtain \eqref{T:crit}. 
 
 The total number of clusters is sub-extensive in the final jammed state:
\begin{equation}
\label{I:critical}
\langle \mathcal{C} \rangle \simeq Nc(T) \sim N T^{-1/5} \sim N^{4/5}
\end{equation}
as was announced in \eqref{clusters:crit}. The mass distribution in the jammed state is unknown. An uncontrolled approximation of the jammed mass distribution is given by the scaling form \eqref{ck:scaling} specialized to time $T$:
\begin{equation}
\label{Ck:crit}
C_k \sim \frac{k-1}{N^{3/5}}\,\exp\!\left[-\frac{k^2}{N^{2/5}}\right]
\end{equation}
This form correctly predicts the typical island mass and the density of such islands:
\begin{equation}
\label{Ck:crit-scaling}
k_\text{typ} \sim N^{1/5}, \qquad C_{k_\text{typ}}\sim N^{-2/5}
\end{equation}
The linear growth of the final mass distribution, $C_k \propto k$, is expected to hold when $1\ll k\ll k_\text{typ} \sim N^{1/5}$. We put the factor $k-1$ into \eqref{Ck:crit} to emphasize that $C_1=0$ by the definition of the jammed state. 

Jammed states have peculiar characteristics in the critical regime: Clusters are large [cf. \eqref{Ck:crit-scaling}], and the number of clusters \eqref{I:critical} is sub-extensive. Fluctuations play a decisive role in the formation and properties of such jammed states. Similar states have been detected, again in the critical regimes, in addition-shattering processes \cite{Wendy21}. They have been called \cite{Wendy21} supercluster states to emphasize that clusters are predominantly large. Supercluster states appear inevitable in finite systems whose infinite-size versions admit jamming and steady-state regimes.

\subsection{Quasi-stationary state regime:  $p < \frac{1}{2}$}
\label{subsec:FS-SS}

In a finite system, the `steady' state is quasi-stationary. If  $N\gg 1$, the system spends an astronomically large time near that state but eventually falls into a jammed state. The average lifetime is exponential in system size:
\begin{equation}
\label{T:SS}
T  \asymp e^{A(p)N}
\end{equation}

Here $\asymp$ means an asymptotic equality of logarithms, i.e., \eqref{T:SS} is the shorthand for the assertion
\begin{equation}
\label{log-T:SS}
\lim_{N\to\infty} N^{-1} \ln T =A(p)
\end{equation}

\begin{figure}[h]
    \centering
    \includegraphics[width=9cm]{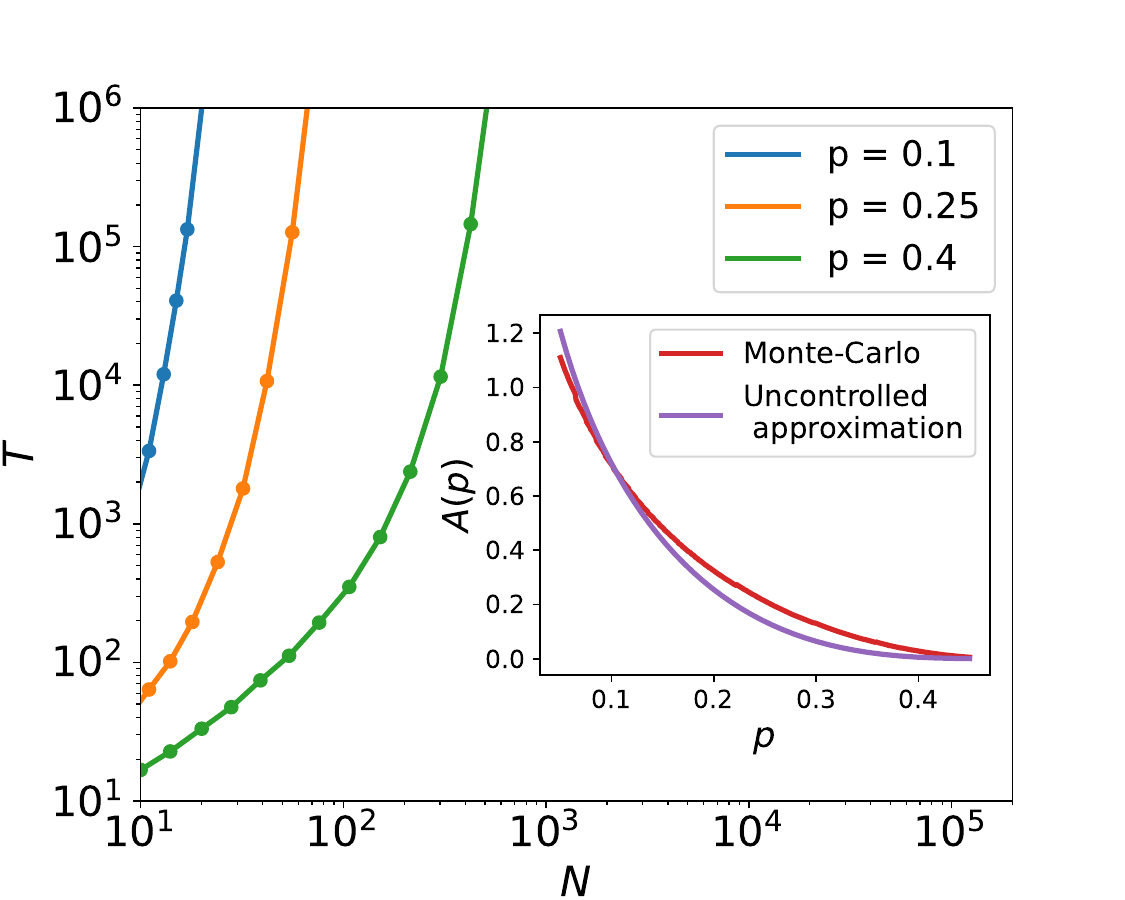}
\caption{For each $N$, the average lifetime is obtained by averaging $10^{3}$ Monte Carlo simulations. An exponential growth is observed, and the amplitude $A(p)$ appearing in \eqref{T:SS} is extracted from the numerical data. The inset shows this amplitude together with an uncontrolled approximation \eqref{A-bar} for the amplitude. An approximation gives a qualitatively correct dependence of the amplitude $A(p)$ on $p$.  }
\label{Fig:A_p}
\end{figure}

The exponential factor $A(p)$ seems to be a complicated nonlinear function of $p < 1/2$ (see Fig. \ref{Fig:A_p}).

Astronomically large adsorption times resembling \eqref{T:SS} arise in population dynamics where they are known as extinction times. Population sizes tend to stay near attracting fixed point of rate equations, but extinction eventually happens after a rare giant fluctuation. Wentzel-Kramers-Brillouin (WKB) technique is a powerful toolbox for finding the controlling exponential behavior \eqref{T:SS}. A WKB theory for stochastic classical systems with continuous phase space is a popular subject \cite{Freidlin}. In the present case, the cluster masses and the numbers of clusters are naturally quantized, and appropriate WKB approximations reflect this feature. A dissipative variant of the WKB was pioneered in \cite{Kubo,Gang,WKB-Peters,WKB-Dykman,Meerson10a}; see \cite{WKB-Kamenev,WKB-KS,Meerson06} for other WKB approaches. 

Single-population models admit analytical treatment \cite{Meerson17}. In such situations, it is often possible to compute the amplitude like $A(p)$ in Eq.~\eqref{T:SS}. Systems with two interacting populations are generally intractable analytically \cite{Meerson17}, albeit the WKB approach tremendously simplifies the analysis leading to a dynamical system with two degrees of freedom (see, e.g., \cite{Meerson08,Dykman08,Meerson12}). The number of interacting populations (island species) in our AC process diverges with $N$, see Sec.~\ref{subsec:species}. Hence an analytical determination of $A(p)$ and the densities in the final jammed state appear impossible. In Appendix~\ref{ap:jammed}, we employ physically appealing yet uncontrolled approximations that lead to qualitatively reasonable predictions.

\subsection{Lifetime distribution}
\label{subsec:time}

The lifetime $\mathcal{T}$ fluctuates from realization to realization. The average lifetime $T=\langle \mathcal{T}\rangle$ is the simplest characteristic of the random variable $\mathcal{T}$. We now argue that in the interesting case of large systems, $N\gg 1$, the average lifetime encodes the chief features of the random variable $\mathcal{T}$ in the jamming and steady-state regimes, $p \ne 1/2$. 

In the jamming regime, we extend the heuristic argument leading to \eqref{T+} and obtain $\mathcal{T}= T + \mathfrak{t}$ with random $\mathfrak{t}=O(1)$. If true,  the variance $V=\langle \mathcal{T}^2\rangle-T^2$ remains finite. Even if the variance grows with size, perhaps logarithmically similar to the average, $V\sim \ln N$, the random variable $\mathcal{T}$ appears self-averaging. This assertion means that deviations from the average are asymptotically negligible compared to the average: 
\begin{equation}
\label{VT:J}
\lim_{N\to\infty}\frac{\sqrt{V}}{T} = 0
\end{equation}
Verifying \eqref{VT:J} numerically is challenging when $p < 1/2$ due to exponential growth of the system lifetime $T$. In the $p \geq 1/2$ regime, our simulations confirm \eqref{VT:J} when $N$ exceeds $10^7$; in our experiments, we are able to utilize $N$ up to $10^8$. For instance, we mentioned $V\sim \ln N$ as a possible large $N$ behavior; if this is true, the ratio in \eqref{VT:J} approaches to zero very slowly, viz., as $(\ln N)^{-1/2}$. 

The lifetime $\mathcal{T}$ of the system in the quasi-stationary regime is a non-self-averaging random variable, i.e., it has a non-trivial distribution. This distribution is (asymptotically) exponential:
\begin{equation}
\label{lifetime:SS}
\text{Prob}[\mathcal{T}=t] = T^{-1} e^{-t/T}
\end{equation}
Proving \eqref{lifetime:SS} for our AC process could be very challenging as the number of interacting populations diverges with $N$. When the number of interacting populations is finite, there is little doubt in the validity of \eqref{lifetime:SS}, see \cite{Meerson17}.  

In the critical regime, the random variable $\mathcal{T}$ is expected to be non-self-averaging, with the lifetime distribution acquiring a scaling form 
\begin{equation}
\label{lifetime:C}
\text{Prob}[\mathcal{T}=t] = T^{-1} P(t/T)
\end{equation}
when $N\gg 1$. The unknown scaled distribution $P(x)$ is probably non-trivial, different from the exponential distribution \eqref{lifetime:SS} in the quasi-stationary regime.

\subsection{Distinct island species and total number of islands}
\label{subsec:species}

The number $\mathcal{D}$ of different island species in the jammed state is 
\begin{equation}
\label{D:def}
\mathcal{D} = \#\{k: \mathcal{C}_k\geq 1\}
\end{equation}
where $\mathcal{C}_k$ is the number of islands of mass $k$. We guess that the average number of different island species exhibits the following growth with $N$ 
\begin{equation}
\label{D:asymp}
D  = \langle \mathcal{D}\rangle \simeq
\begin{cases}
\frac{\ln N}{\ln(\ln N)}              & p > \frac{1}{2}\\
D_c\, N^{1/5}\sqrt{\ln N}         & p = p_c = \frac{1}{2}\\
D_-(p)\,\ln N                            & p < \frac{1}{2}
\end{cases}
\end{equation}

The evidence is favor of \eqref{D:asymp} is rather slim. Let us begin with the jamming regime. In the extreme case of pure addition process, $p=1$, the jammed densities are given by \eqref{ck-final}. The criterion 
\begin{equation}
\label{crit:different}
N C_D\sim 1
\end{equation}
and the factorial decay \eqref{ck-final} yield
\begin{equation}
D \simeq \frac{\ln N}{\ln(\ln N)-1} 
\end{equation}
where we also used the asymptotic $\ln k!\simeq k(\ln k-1)$ implied by the Stirling formula. The analog of \eqref{ck-final} in the jamming regime $1/2 < p \leq 1$ is in principle contained in the exact expression \eqref{Lap:sol} for the Laplace transform.  Here we just extract from \eqref{Lap:sol} the large $s$ asymptotic $\widehat{c}_k\simeq p^{k-1}/s^k$ implying the small $\tau$ asymptotic
\begin{equation}
\label{Ck-small-jammed}
C_k \simeq \frac{(p\tau)^{k-1}}{(k-1)!}
\end{equation}
The jammed regime is formed at $\tau=\tau_\text{max}$ which  is not small, but still using \eqref{Ck-small-jammed} and \eqref{crit:different} we obtain
\begin{equation}
\label{D+}
D \simeq \frac{\ln N}{\ln(\ln N)-1-\ln(p\tau_\text{max})} 
\end{equation}
The dependence on $p$ disappears only when $N$ becomes astronomically large, viz., $\ln(\ln N)\gg 1$. 

In the critical regime, we use the mass distribution \eqref{Ck:crit} in the jammed state and the criterion \eqref{crit:different} to get the result quoted in \eqref{D:asymp}. In the quasi-stationary regime, the final mass distribution is expected to have an exponential large mass tail:
\begin{equation}
\label{Ck-QS}
C_k \asymp e^{-k/D_-(p)}
\end{equation}
Equations \eqref{crit:different} and \eqref{Ck-QS} give the result quoted in \eqref{D:asymp}. If the tail is the same as in the steady-state, \eqref{ckc:final}, 
\begin{equation}
\label{D-minus}
D_-(p) = \frac{1}{\ln(1/p-1)}
\end{equation}

\begin{figure}[h]
    \centering
    \includegraphics[width=9cm]{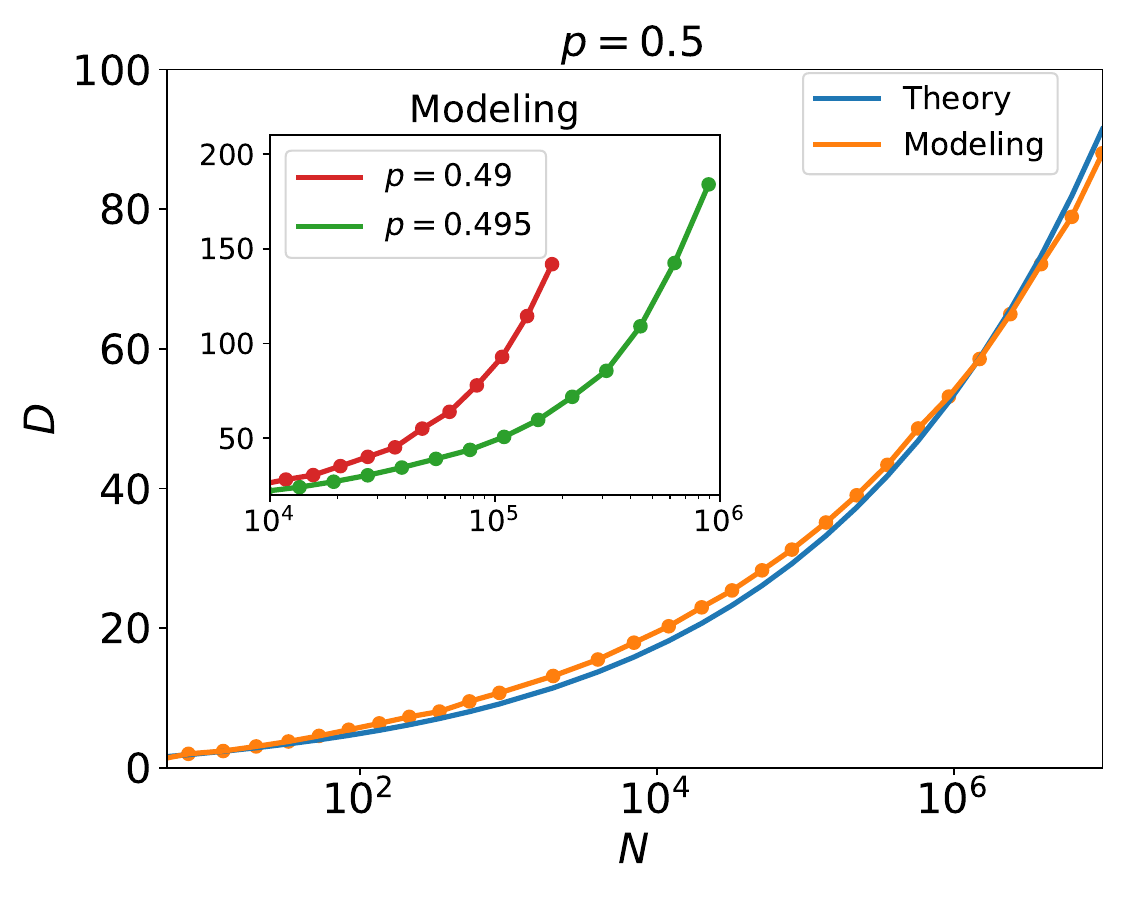}
\caption{The average number of different island species obtained by averaging $10^3$ Monte Carlo runs for each value of $N$. When $p=1/2$, simulation results are in good agreement with the theoretical prediction \eqref{D:asymp}. For $p < 1/2$, we could not verify the theory (see the inset): The system size seems insufficient for reaching the $\mathcal{D} \simeq D_{-}(p)\, \ln N$ asymptotic.}
\label{Fig:Dist_islands}
\end{figure}

\begin{figure}[h]
    \centering
    \includegraphics[width=9cm]{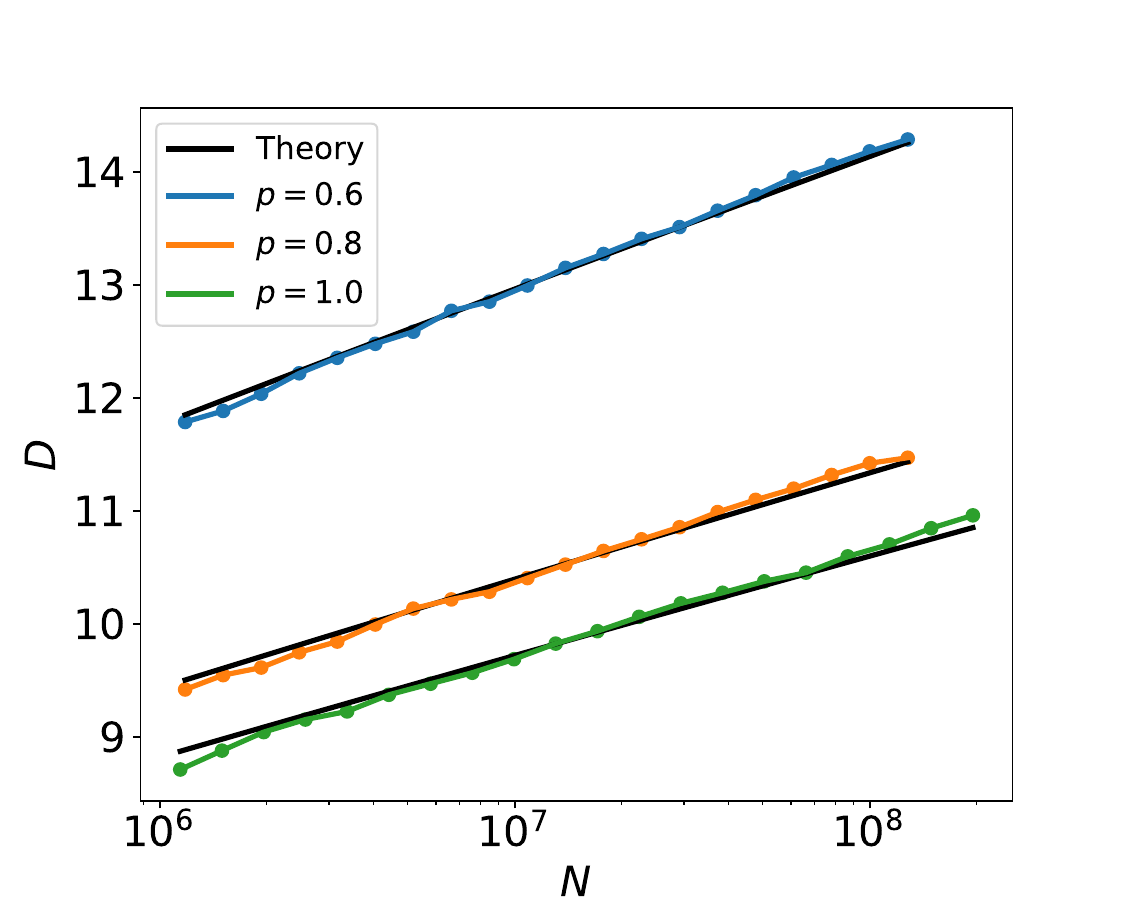}
\caption{The average number of different island species obtained by averaging $10^3$ Monte Carlo runs for each value of $N$ in the jammed state for several values of $p>1/2$. Numerical results agree with \eqref{D:asymp} and demonstrate the dependence of $D_{+}(p)$ on $p$. The number of island species grows when $p$ decreases. }
    \label{Fig:D+}
\end{figure}

Simulation results (see Figs.~\ref{Fig:Dist_islands}--\ref{Fig:D+}) qualitatively agree with theoretical predictions when $p \geq 1/2$. In the quasi-stationary regime, $p<1/2$, we observe a faster than logarithmic growth (see the inset in Fig.~\ref{Fig:Dist_islands}). Numerical experiments are extremely time-consuming when $p<1/2$, apart from the situation when $\frac{1}{2}-p\ll 1$. This is close to the critical regime where the growth is indeed faster than logarithmic. 

\begin{figure}[h]
    \centering
    \includegraphics[width=9cm]{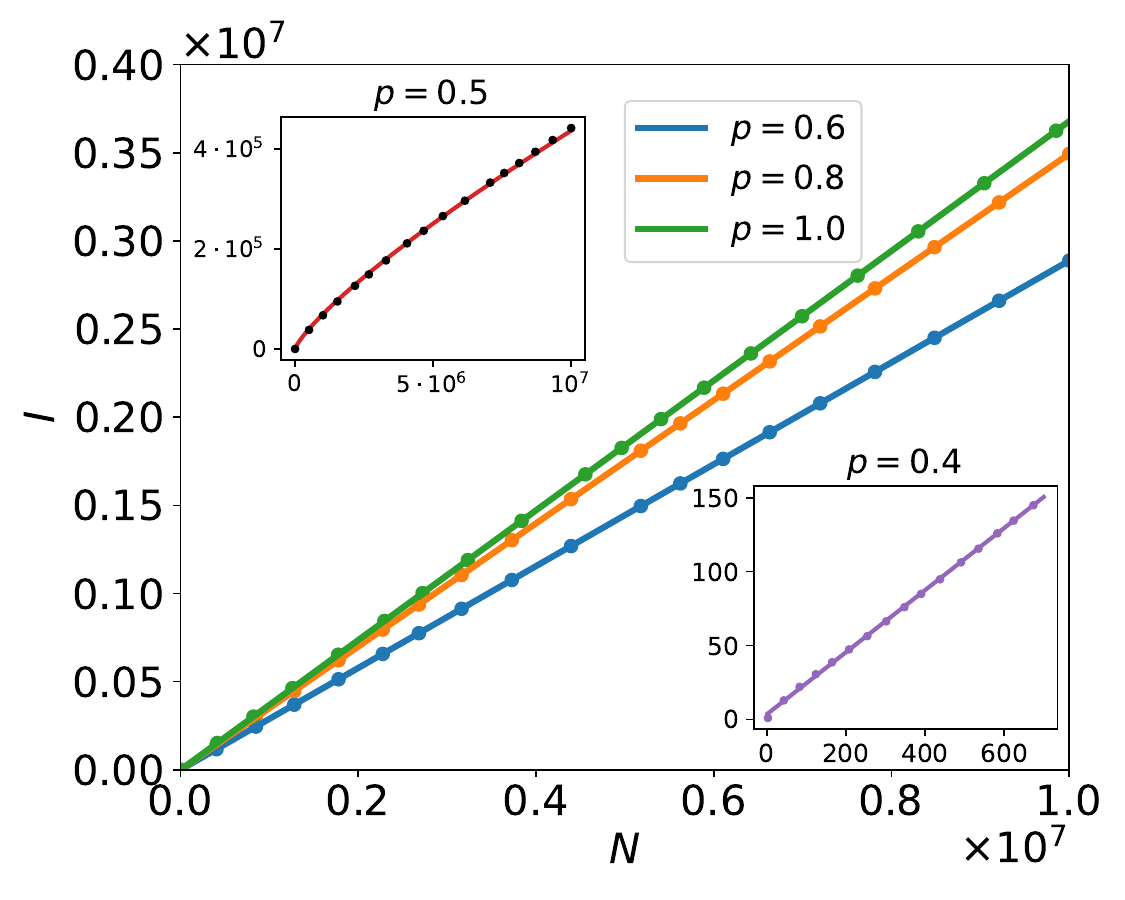}
\caption{The average total number of clusters. 
Numerical simulations (lines) of the AC process agree with theory \eqref{I:asymp} shown by dots. For $p < 1/2$, the simulations also agree with the analysis but become very time-consuming for $N > 10^{3}$ due to the exponential growth of the system lifetime. Simulations indicate that both $A_+(p)$ and $A_{-}(p)$ depend on $p$.}
\label{Fig:Total_particles}
\end{figure}

The total number of clusters $\mathcal{C}$ in the final state also grows with system size. The average growth is
\begin{equation}
\label{I:asymp}
I = \langle \mathcal{C}\rangle \simeq
\begin{cases}
A_+(p)\, N              & p > \frac{1}{2}\\
A_c\,N^{4/5}          & p = p_c = \frac{1}{2}\\
A_-(p)\, N              & p < \frac{1}{2}
\end{cases}
\end{equation}
and it agrees with our numerical observations for all values of $0<p<1$ (see Fig. \ref{Fig:Total_particles}). Combining \eqref{I:asymp} and 
\begin{equation}
\label{I:jamming}
I = NC
\end{equation}
we see that in the thermodynamic $N\to\infty$ limit, the final cluster density vanishes in the supercluster state, and remains positive otherwise: $C=A_+(p)$ for $p > \frac{1}{2}$ and $C=A_-(p)$ for $p < \frac{1}{2}$.

Fluctuations around the average are relatively small. Indeed, using van Kampen expansion we write 
\begin{align}
\label{I:def}
\mathcal{C}(t) = N c(t) + \sqrt{N} \eta(t)
\end{align}
In terms of the modified time, the evolution span is finite, $\tau\leq  \tau_\text{max}$. The random variable $\eta(\tau_\text{max})$ is expected to remain finite, and hence the fluctuations around $I=NC$ are of the order of $\sqrt{N}$. Thus $\mathcal{C}$ is a self-averaging random quantity and \eqref{I:jamming} fixes the amplitude
\begin{equation}
A_+(p) = C_\infty(p) 
\end{equation}
This amplitude admits explicit expressions  in extreme situations close to the critical regime $\big(p\downarrow 1/2\big)$, and to the pure addition process ($p\uparrow 1$):
\begin{equation}
\label{A-plus}
A_+(p) \simeq 
\begin{cases}
2\sqrt{\frac{2p-1}{\pi}}                   &0 < 2p - 1 \ll 1\\
e^{-1}\left[1+\frac{p-1}{6}\right]      &0< 1-p \ll 1
\end{cases}
\end{equation}

We have argued that the total number of islands in the jamming regime is sub-extensive in the final jammed state, see \eqref{I:critical}.  It would be interesting to investigate numerically the nature of the random quantity $\mathcal{C}$ characterizing the supercluster state at the critical regime. It is  probably a non-self-averaging random quantity characterized by non-trivial scaled distribution
\begin{equation}
\label{SCS}
\text{Prob}[\mathcal{C}=\mathfrak{C}] = N^{-4/5} F(\mathfrak{C}/N^{4/5})
\end{equation}

In the steady state regime before giant fluctuation we have $NC$ islands and  $NC_1$  monomers with  [see Eq.~\eqref{ckc:final}]
\begin{equation}
\label{CC1}
C = \frac{1-2p}{1-p}, \quad C_1 = C^2
\end{equation}
In a giant fluctuation almost all collision events are additions. Schematically
\begin{equation}
\mathbb{A}+\mathbb{I}\to \mathbb{I}, \quad \mathbb{A}+\mathbb{A}\to\mathbb{I}
\end{equation}
where we have disregarded island masses. Thus we arrive at the bounds $C\leq  A_-(p)\leq  C + C_1/2$. Using \eqref{CC1} we re-write these bounds as
\begin{equation}
\frac{1-2p}{1-p} \leq A_-(p)\leq \frac{(1-2p)\big(\frac{3}{2}-2p\big)}{(1-p)^2}
\end{equation}

\section{Fluctuations in the critical regime}
\label{sec:F}

Denote by $\mathcal{C}_k(t)$ the total number of clusters of size $k$. At any time, the state of the finite system is represented by configuration $\{\mathcal{C}_1(t), \mathcal{C}_2(t), \ldots, \mathcal{C}_N(t)\}$. All $\mathcal{C}_k(t)$ are non-negative integers satisfying the constraint
\begin{equation}
\label{mass:N}
\sum_{j=1}^N j \mathcal{C}_j(t) = N
\end{equation}
implied by mass conservation. The initial configuration is $\{N, 0, \ldots, 0\}$.

In a reaction event, the configuration $\{\mathcal{C}_1, \mathcal{C}_2, \ldots, \mathcal{C}_N\}$ transforms into one of the following configurations
\begin{subequations}
\label{Exact}
\begin{align}
\label{channel:m}
& (\mathcal{C}_1-2, \mathcal{C}_2+1)     ~\quad \qquad \qquad    \mathcal{C}_1(\mathcal{C}_1-1)/(2N) \\
\label{channel:add}
&(\mathcal{C}_1-1,\mathcal{C}_k-1,\mathcal{C}_{k+1}+1)              \quad \mathcal{C}_1 \mathcal{C}_k/(2N)  \\
\label{channel:d}
& (\mathcal{C}_1+2, \mathcal{C}_2-1)     ~\quad \qquad \qquad     \mathcal{C}_1\mathcal{C}_2/(2N)    \\
\label{channel:chip} 
&(\mathcal{C}_1+1,  \mathcal{C}_{k-1}+1  ,   \mathcal{C}_k-1)       \quad \mathcal{C}_1 \mathcal{C}_k/(2N)
\end{align}
\end{subequations}
 in the critical regime. To avoid cluttering, we only show components of an evolved configuration that differ from the corresponding components of the original configuration; the rates of reaction channels are also shown in Eqs.~\eqref{Exact}. The last reaction channel \eqref{channel:chip} describes the chipping of clusters with $k\geq 3$; the chipping process with $k=2$ is represented by \eqref{channel:d}.

Using Eqs.~\eqref{Exact} we deduce equations for the averages
\begin{subequations}
\label{Av-eqs}
\begin{align}
\label{av-1-eq}
&2N\,\frac{d \langle \mathcal{C}_1\rangle}{dt} = - 2 \langle \mathcal{C}_1(\mathcal{C}_1-1)\rangle +  \langle \mathcal{C}_1 \mathcal{C}_2\rangle \\
\label{av-2-eq}
&2N\,\frac{d \langle \mathcal{C}_2\rangle}{dt} =  \langle \mathcal{C}_1(\mathcal{C}_1-1)\rangle -  2\langle \mathcal{C}_1 \mathcal{C}_2\rangle 
                                                                             +\langle \mathcal{C}_1 \mathcal{C}_3\rangle \\
\label{av-k-eq}
&2N\,\frac{d \langle \mathcal{C}_k\rangle}{dt} =  \langle \mathcal{C}_1\mathcal{C}_{k-1}\rangle -  2\langle \mathcal{C}_1 \mathcal{C}_k\rangle 
                                                                             +\langle \mathcal{C}_1 \mathcal{C}_{k+1}\rangle 
\end{align}
\end{subequations}

Using the van Kampen expansion \eqref{M:def} for $\mathcal{C}_1$ and similar expansions 
\begin{align}
\label{Ck:def}
\mathcal{C}_k(t) = N c_k(t) + \sqrt{N} \xi_k(t)
\end{align}
we compute
\begin{equation}
\label{CCCk}
\begin{split}
\langle \mathcal{C}_1\rangle  &= N c_1 + \sqrt{N}\,\langle \xi_1\rangle   \\
\langle \mathcal{C}_1^2\rangle &=N^2 c_1^2 + 2 N^{3/2} c_1\, \langle \xi_1\rangle + N \langle \xi_1^2\rangle  \\
\langle \mathcal{C}_1  \mathcal{C}_k\rangle  &= N^2 c_1 c_k+ N^{3/2}\,[c_1\langle \xi_k\rangle+ c_k\langle \xi_1\rangle] \\
&+ N \langle \xi_1 \xi_k\rangle
\end{split}
\end{equation}
Plugging these expansions into \eqref{av-1-eq} and equating the leading $O(N^2)$ terms we recover the rate equation for the density of monomers. Equating  sub-leading $O(N^{3/2})$ terms we arrive at
\begin{subequations}
\label{xi-123}
\begin{equation}
\label{xi-1}
2\frac{d \langle \xi_1\rangle}{dt} = -4 c_1 \langle \xi_1\rangle + c_1\langle \xi_2\rangle+ c_2\langle \xi_1\rangle
\end{equation}
Similarly from \eqref{av-2-eq} we deduce
\begin{eqnarray}
\label{ap:xi-2}
2\frac{d \langle \xi_2\rangle}{dt} &=& 2 c_1 \langle \xi_1\rangle -2[c_1\langle \xi_2\rangle+ c_2\langle \xi_1\rangle] \nonumber\\
&+& c_1\langle \xi_3\rangle+ c_3\langle \xi_1\rangle
\end{eqnarray}
and Eqs.~\eqref{av-k-eq} give
\begin{eqnarray}
\label{xi-3}
2\frac{d \langle \xi_k\rangle}{dt} &=& c_1 [\langle \xi_{k-1}\rangle -2\langle \xi_k\rangle+ \langle \xi_{k+1}\rangle] \nonumber\\
&+&\langle \xi_k\rangle [c_{k-1}-2c_k+c_{k+1}]
\end{eqnarray}
\end{subequations}
for $k\geq 3$.  The initial state is deterministic: $\xi_k(0)=0$ for all $k\geq 1$. An infinite homogeneous system \eqref{xi-123} of linear equations with initial condition $\langle \xi_k(0)\rangle = 0$ has a trivial vanishing solution:
\begin{equation}
\langle \xi_k\rangle\equiv 0
\end{equation}
Hence the first and  second-order cumulants become
\begin{equation}
\label{ij-av}
\langle \mathcal{C}_j\rangle   = N c_j, \quad 
\langle \mathcal{C}_i  \mathcal{C}_j\rangle_c =  N W_{i j}
\end{equation}
where $\langle \mathcal{C}_i  \mathcal{C}_j\rangle_c=\langle \mathcal{C}_i  \mathcal{C}_j\rangle- \langle \mathcal{C}_i\rangle \langle \mathcal{C}_j\rangle$
and we shortly write $W_{i j}=\langle \xi_i \xi_j\rangle$.

To derive the evolution equation for $W_{11} = \langle \xi_1^2\rangle$ we first notice that $\langle \mathcal{C}_1^2\rangle$ obeys
\begin{eqnarray}
\label{MM-eq}
2N\,\frac{d \langle \mathcal{C}_1^2\rangle}{dt} &=& \langle \mathcal{C}_1(\mathcal{C}_1-1)(-4\mathcal{C}_1+4)\rangle \nonumber \\
&+&  \sum_{k\geq 2} \langle \mathcal{C}_1 \mathcal{C}_k(-2\mathcal{C}_1+1)\rangle  \nonumber \\
&+&  \langle \mathcal{C}_1 \mathcal{C}_2 (4\mathcal{C}_1+4)\rangle  \nonumber \\
&+&  \sum_{k\geq 3} \langle \mathcal{C}_1 \mathcal{C}_k(2\mathcal{C}_1+1)\rangle
\end{eqnarray}
Each term on the right-hand side (top to bottom) corresponds to the corresponding reaction channel in \eqref{Exact}. Massaging the right-hand side of \eqref{MM-eq} we obtain
 \begin{eqnarray}
\label{MM-short}
2N\,\frac{d \langle \mathcal{C}_1^2\rangle}{dt} &=& -4\langle \mathcal{C}_1(\mathcal{C}_1-1)^2\rangle \nonumber \\
&+&  2\sum_{k\geq 2} \langle \mathcal{C}_1 \mathcal{C}_k\rangle +  \langle \mathcal{C}_1 \mathcal{C}_2 (2\mathcal{C}_1+3)\rangle
\end{eqnarray}
Combining \eqref{av-1-eq} and \eqref{MM-eq} we obtain
 \begin{eqnarray}
\label{M-var}
N\,\frac{d \langle \mathcal{C}_1^2\rangle_c}{dt} &=& 
2[\langle \mathcal{C}_1^2\rangle \langle \mathcal{C}_1\rangle-\langle \mathcal{C}_1^3\rangle] + \tfrac{3}{2}\langle \mathcal{C}_1  \mathcal{C}_2\rangle \nonumber \\
&+&\langle \mathcal{C}_1^2  \mathcal{C}_2\rangle-\langle \mathcal{C}_1\rangle \langle \mathcal{C}_1  \mathcal{C}_2\rangle + \sum_{k\geq 2} \langle \mathcal{C}_1 \mathcal{C}_k\rangle  \nonumber \\
&+&4\langle \mathcal{C}_1^2 \rangle_c - 2 \langle \mathcal{C}_1\rangle
\end{eqnarray}
We now compute the leading behavior  of the third order moments
\begin{equation}
\label{112-av}
\begin{split}
&  \langle \mathcal{C}_1^3\rangle- \langle \mathcal{C}_1^2\rangle \langle \mathcal{C}_1\rangle = 2N^2 c_1 W_{11}  \\
& \langle \mathcal{C}_1^2  \mathcal{C}_2\rangle -\langle \mathcal{C}_1\rangle \langle \mathcal{C}_1  \mathcal{C}_2\rangle =  N^2[c_1 W_{12}  +  c_2 W_{11}]  
\end{split}
\end{equation}
Inserting \eqref{ij-av} and  \eqref{112-av}  into \eqref{M-var}, keeping the leading $O(N^2)$ terms and using the modified time variable gives
\begin{equation}
\label{V-eq}
\frac{d W_{11}}{d \tau} =W_{12}-\left(4-\frac{c_2}{c_1}\right)W_{11}+ c-c_1 + \frac{3}{2}\,c_2
\end{equation}

We thus also need to derive the evolution equation for $W_{12} = \langle \xi_1 \xi_2\rangle$.  A lengthy calculation yields
\begin{eqnarray}
\label{C12-eq}
2N\,\frac{d \langle \mathcal{C}_1 \mathcal{C}_2\rangle}{dt} &=& \langle \mathcal{C}_1^3\rangle - 4 \langle \mathcal{C}_1^2 \mathcal{C}_2\rangle
+  \langle \mathcal{C}_1 \mathcal{C}_2^2\rangle +  \langle \mathcal{C}_1^2 \mathcal{C}_3\rangle \nonumber \\
&+&  \langle \mathcal{C}_1 (\mathcal{C}_2 + \mathcal{C}_3)\rangle  -3 \langle \mathcal{C}_1^2\rangle +  2 \langle \mathcal{C}_1\rangle
\end{eqnarray}
which in conjunction with Eqs.~\eqref{av-1-eq}--\eqref{av-2-eq} lead to
\begin{eqnarray}
\label{C12-eq-c}
2N\,\frac{d \langle \mathcal{C}_1 \mathcal{C}_2\rangle_c}{dt} &=& \langle \mathcal{C}_1^3\rangle - \langle \mathcal{C}_1^2\rangle \langle \mathcal{C}_1\rangle \nonumber \\
&-& 2[\langle \mathcal{C}_1^2 \mathcal{C}_2\rangle-\langle \mathcal{C}_1 \mathcal{C}_2\rangle \langle \mathcal{C}_1\rangle] \nonumber \\
&-& 2[\langle \mathcal{C}_1^2 \mathcal{C}_2\rangle -  \langle \mathcal{C}_1^2\rangle \langle \mathcal{C}_2\rangle]  \nonumber \\
&+&  \langle \mathcal{C}_1 \mathcal{C}_2^2\rangle -  \langle \mathcal{C}_1 \mathcal{C}_2\rangle  \langle \mathcal{C}_2\rangle  \nonumber \\
&+&  \langle \mathcal{C}_1^2 \mathcal{C}_3\rangle - \langle \mathcal{C}_1 \mathcal{C}_3\rangle  \langle \mathcal{C}_1\rangle  \nonumber \\
&+&  \langle \mathcal{C}_1 ( \mathcal{C}_3 - \mathcal{C}_2)\rangle  -2 \langle \mathcal{C}_1^2\rangle 
\end{eqnarray}
where we have dropped the  sub-leading $2 \langle \mathcal{C}_1\rangle$  term. We already know the leading behavior of the terms in the top to lines, see \eqref{112-av}. Similarly we compute
\begin{equation}
\label{123-av}
\begin{split}
& \langle \mathcal{C}_1^2 \mathcal{C}_2\rangle -  \langle \mathcal{C}_1^2\rangle \langle \mathcal{C}_2\rangle =  2  N^2 c_1 W_{12} \\
& \langle \mathcal{C}_1 \mathcal{C}_2^2\rangle -  \langle \mathcal{C}_1 \mathcal{C}_2\rangle  \langle \mathcal{C}_2\rangle =  
N^2 [c_2 W_{12}  +  c_1 W_{22}]\\
& \langle \mathcal{C}_1^2 \mathcal{C}_3\rangle - \langle \mathcal{C}_1 \mathcal{C}_3\rangle  \langle \mathcal{C}_1\rangle = 
N^2[c_1 W_{13}  +  c_3 W_{11}] 
\end{split}
\end{equation}
from which
\begin{eqnarray}
\label{W-eq}
\frac{d W_{12}}{d \tau} &=& \left(2-\frac{2c_2}{c_1}+\frac{c_3}{c_1}\right)W_{11}- \left(6-\frac{c_2}{c_1}\right)W_{12} \nonumber\\
&+&W_{22}+W_{13}+c_3-c_2-2c_1
\end{eqnarray}

Thus, we must derive equations for $W_{22}$ and $W_{13}$. The good news is that Eqs.~\eqref{V-eq}, \eqref{W-eq}, and equations for other cumulants do not involve higher cumulants. The bad news is that equations are hierarchical and hence seem intractable. In the long time limit $c_j/c_1\simeq j$ and $c_k \ll c$, so Eqs.~\eqref{V-eq}, \eqref{W-eq} simplify to  
\begin{subequations}
\begin{align}
\label{W11}
\frac{d W_{11}}{d \tau} &=W_{12}-2 W_{11}+ c\\
\label{W12}
\frac{d W_{12}}{d \tau} &=W_{11}-4 W_{12}+ W_{22}+W_{13}
\end{align}
\end{subequations}
In the long time limit equations for $W_{ij}$ are similar to \eqref{W12}, namely, the right-hand sides are linear combinations of cumulants, and only 
\eqref{W11}  additionally contains $c$. It seems that $W_{ij}=A_{ij} c$ in the long time limit. The amplitudes  $A_{ij}$ are unknown, but $W_{11}=\langle \xi_1^2\rangle  \sim c$ already suffices to establish \eqref{T:crit} as we have shown in Sec.~\ref{subsec:FS-C}.

\section{Addition and chipping processes with proportional rates}
\label{sec:prop}

The AC processes with proportional rates
\begin{equation}
\label{AC-prop}
\mathrm{C}_k = \lambda \mathrm{A}_k \quad \text{when} \quad k\geq 2
\end{equation}
behave similarly to the processes with mass-independent rates analyzed in Secs.~\ref{sec:ACP}--\ref{sec:F}: The outcome depends on which of the two processes is more potent, i.e., whether $\lambda$ smaller, equal, or larger than unity.

Specifically, we looked at models with algebraic rates
\begin{equation}
\label{Ak-alg}
\mathrm{A}_k = k^a
\end{equation}
satisfying \eqref{AC-prop}. The two most tractable AC processes of the type \eqref{AC-prop} are the model with $a=0$ (mass-independent rates) and $a=1$ (rates proportional to the mass of the cluster participating in a collision). The latter processes arise in several applications \cite{economy-Slava,migration,BK-exchange} providing extra motivation to extend our theoretical analysis of the AC processes with mass-independent rates ($a=0$) for the model with $a=1$. In this situation 
\begin{equation}
\label{A-rate-lin}
\mathrm{A}_k = k
\end{equation}
and then \eqref{AC-prop} becomes 
\begin{equation}
\label{C-rate-lin}
\mathrm{C}_k = \lambda k \quad \text{when} \quad k\geq 2
\end{equation}

A finite system gets jammed with probability one. The time to reach a jammed state scales according to
\begin{equation}
\label{lifetime-linear}
T \propto
\begin{cases}
\ln N               &  \lambda < 1\\
N^{3/4}           & \lambda = 1\\
e^{D(\lambda) N}      & \lambda > 1
\end{cases}
\end{equation}
for the AC processes with rates \eqref{A-rate-lin}--\eqref{C-rate-lin}. The critical regime is again particularly interesting as the final supercluster state is non-extensive. The final number of clusters and the typical cluster mass scale according to
\begin{equation}
\label{mass:crit-linear}
\mathcal{C} \sim N^\frac{3}{4}\,, \qquad k_\text{typ} \sim  N^\frac{1}{4} 
\end{equation}
in the supercluster state. The details of derivation of \eqref{lifetime-linear}--\eqref{mass:crit-linear} and other results are relegated to Appendix~\ref{ap: AC_linear}.

For the AC processes with algebraic rates \eqref{AC-prop}--\eqref{Ak-alg}, we mostly looked at supercluster states. In Appendix~\ref{ap: AC_alg} we estimate the time to reach the supercluster state 
\begin{equation}
\label{T-a}
T \sim N^\frac{5-2a}{5-a}
\end{equation}
and argue that the final number of clusters and the typical cluster mass scale according to 
\begin{equation}
\label{number-mass:a}
\mathcal{C} \sim N^\frac{4-a}{5-a}\,, \qquad k_\text{typ} \sim  N^\frac{1}{5-a} 
\end{equation}

\section{Discussion}
\label{sec:disc}

We analyzed addition and chipping  (AC) processes with proportional reaction rates \eqref{AC-prop}. We primarily focused on the ultimate fate of finite systems. Since both addition and chipping processes are driven by collisions with monomers, any finite system eventually reaches a jammed state without monomers where evolution ceases. The route to the final state and its composition greatly depend on which of the two processes prevails. 

If addition prevails ($\lambda<1$), the system quickly relaxes to a jammed state close to the jammed state of an infinite system. When chipping prevails ($\lambda>1$), an infinite system relaxes to a steady state with a positive density of monomers. For a long time, a large finite system remains in a quasi-stationary state with densities fluctuating around the densities of the steady state of an infinite system. Eventually, monomers disappear in a huge fluctuation. The average lifetime of the quasi-stationary state scales exponentially with the total mass 
\begin{equation}
\label{T:QS}
T  \asymp e^{A(\lambda, a)N}
\end{equation}

These phenomena seem quite general. For instance, similar behaviors have been numerically observed for collision-controlled aggregation-shattering systems \cite{kalinov2022direct}. In this case, the finite system falls out from the limiting cycle predicted for the infinite case \cite{Sergei17,Sergei18}.

Computing the amplitude $A(\lambda, a)$ for the class of AC processes with algebraic rates \eqref{AC-prop}--\eqref{Ak-alg} is an outstanding challenge. WKB approaches (see \cite{Meerson17} for review) have been successfully applied to the determination of controlling exponential factors similar to \eqref{T:QS}. However, these approaches are potent only in the case of a few interacting populations, and even in those situations, an analytical treatment tends to work only for a single self-interacting population. For the AC processes the number of interacting cluster species diverges logarithmically with mass. Thus we do not know how to determine the amplitude $A(\lambda, a)$ in \eqref{T:QS} and the composition of the final jammed state when $\lambda>1$. 

The critical regime, $\lambda=1$, is the most interesting already for an infinite system. In two particularly tractable critical AC processes, namely for the model with mass-independent rates ($a=0$), and for the model with linear in mass rates ($a=1$), the full time-dependent solutions for the mass distribution are available [Sec.~\ref{subsec:critical} and  Appendix~\ref{ap: AC_linear}]. For the critical AC processes with algebraic rates \eqref{AC-prop}--\eqref{Ak-alg}, the mass distribution acquires a scaling form in the large time limit [Appendix~\ref{subsec: AC_alg}], and the scaled mass distribution is known for arbitrary $a<2$. For finite critical AC processes, the final jammed state known as supercluster state is quite remarkable, e.g., the final number of clusters is non-extensive in the total mass $N$ and the typical cluster mass algebraically diverges with $N$, see \eqref{number-mass:a}. The outcomes also exhibit large fluctuations from realization to realization, a manifestation of the lack of self-averaging. 

Our derivation relies on the van Kampen expansion applicable when $N\gg 1$. We then estimate the lifetime (i.e., the time $T$ when the last monomer disappears) by equating the deterministic part $Nc_1(T)$ and the stochastic part $\sqrt{N}\xi_1(T)$. Namely, we use the criterion
\begin{equation}
\label{criterion}
N c_1(T) \sim \sqrt{N W_{11}(T)}\,, \quad W_{11} = \langle  \xi_1^2\rangle
\end{equation}
Thus we must determine the variance $W_{11}$. Using the van Kampen expansion, we derived an evolution equation for $W_{11}$ that contains $W_{12} = \langle  \xi_1 \xi_2\rangle$. An evolution equation for $W_{12}$ contains  $W_{11}, W_{13}, W_{22}$. Continuing, one arrives at an infinite set of linear equations for  $W_{ij} = \langle  \xi_i \xi_j\rangle$. We have not verified it in full detail, but in the long time limit, all these equations are homogeneous apart from an evolution equation for $W_{11}$ that contains 
\begin{equation}
\label{m-a:def}
m_a = \sum_{k\geq 1}k^a c_k
\end{equation}
Thus the solution of the infinite set of linear equations is $W_{ij}=C_{ij}m_a$ with some numerical factors $C_{ij}$. The precise value of $C_{11}$ is unknown, but it is not necessary as we only seek the scaling dependence of $T$ from $N$. The \eqref{criterion} thus becomes
\begin{equation}
\label{crit:NW}
N  \sim \frac{m_a(T)}{[c_1(T)]^2}
\end{equation}

The non-rigorous ingredient is that the van Kampen expansion used in deriving equations for $W_{ij}$ tacitly assumes that the deterministic part substantially exceeds the stochastic part. Moreover, we insert the deterministic predictions for $m_a$ and $c_1$ into \eqref{crit:NW}. Still, the emerging scaling laws \eqref{T-a}--\eqref{number-mass:a} appear to be exact.

We studied numerically the simplest AC process with mass-independent rates ($a=0$). Simulation results qualitatively agree with our theoretical predictions. The observed quantitative disagreements are not surprising as the true asymptotic behavior of several quantities emerges only when $\ln N\gg 1$. One quantity contains a repeated logarithm [cf. Eqs.~\eqref{D:asymp} and \eqref{D+}], so the true asymptotic formally emerges when $\ln(\ln N)\gg 1$. In simulations, we used a direct but  efficient Monte Carlo algorithm \cite{Os:LRMC}. Implementing special tricks for handling rare event simulations (see, e.g., \cite{dandekar2023monte}) may significantly contribute to probing $\lambda\leq 1$ regimes.

Input of adatoms is crucial in applications in surface science \cite{MBE}. Pure addition processes with input investigated in \cite{BK91,Blackman91} exhibit very different behavior than the same addition processes without input. It would be interesting to investigate the influence of input of monomers in the AC processes \eqref{AC-prop}--\eqref{Ak-alg}. In this infinite system jamming is impossible due to the constant input of monomers. 

We emphasize that we relied on a mean-field approach. An infinite AC process is then described by an infinite set of ordinary differential equations for the densities. To examine the influence of space in the simplest setting, one can consider a point-island model postulating that each cluster occupies a singe lattice site, and monomers hop to neighboring sites while islands are immobile. 

Unfortunately, there is no analytical framework already for the point-island pure addition process. Moreover, even if we disregard the distinction between islands, so that the process is represented by the reaction scheme
\begin{equation}
\mathbb{A}+\mathbb{A}\longrightarrow \mathbb{I}, \quad  \mathbb{A}+\mathbb{I} \longrightarrow \mathbb{I}
\end{equation}
involving just two populations, adatoms and islands, the problem remains analytically intractable. The pure addition process in the point-island setting was numerically studied in one and two dimensions in Ref.~\cite{Sander}. 

The point-island AC process also depends on $p$ and, additionally, on the density $\rho$ and the spatial dimension $d$.  (For concreteness, we consider the process on hyper-cubic lattices $\mathbb{Z}^d$.) The critical probability is $p_c=\frac{1}{2}$ in the mean-field realm, while for the point-island AC process $p_c=p_c(\rho, d)$. It would be interesting to explore the final state in the critical regime in a finite system. The simplest example is a ring with $L$ sites and total mass $N$, so the density is $\rho=N/L$.  

\bigskip\noindent
The work of R.R.D and S.A.M was partly supported by the Moscow Center of Fundamental and Applied Mathematics at INM RAS (grant No.075-15-2022-286).

\appendix

\section{A jammed state in the quasi-stationary regime}
\label{ap:jammed}

Here we estimate a typical lifetime and basic features of a jammed state in the quasi-stationary regime. The system spends a long time in a state close to the steady state of an infinite system where the densities of monomers and clusters are 
\begin{equation}
\label{CC1-0}
C_1 = \left(\frac{1-2p}{1-p}\right)^2\,, \qquad C = \frac{1-2p}{1-p}
\end{equation}
The fastest (in terms of the number of collisions) path to jamming occurs if each collision involves two monomers. The number of collisions is $C_1^2N/2$, and addition occurs in each event with probability. Hence $p^{-C_1^2N/2}$ provides an estimate of a typical lifetime, from which we deduce an uncontrolled approximation 
\begin{equation}
\label{A-bar}
\overline{A} = \frac{1}{2}\left(\frac{1-2p}{1-p}\right)^4\,\ln(1/p)
\end{equation}
qualitatively agreeing with numerical data (see the inset in Fig.~\ref{Fig:A_p}). 


To estimate the densities in the final jammed state we also {\em assume} that each each collision leads to addition. If as above we make an extra assumption that only monomers collide with each other, the final density of dimers increases 
\begin{equation}
\overline{C}_2=C_2+\tfrac{1}{2}C_1 = \frac{(1-2p)^2\,(1+p)}{2(1-p)^3}
\end{equation}
Other densities remain the same: $\overline{C}_k=C_k$ for $k\geq 3$. The total density is therefore
\begin{equation}
\label{C-bar}
\overline{C}=C - \tfrac{1}{2}C_1 = \frac{1-2p}{2(1-p)^2}
\end{equation}

\begin{figure}
\centering
\includegraphics[width=7.77cm]{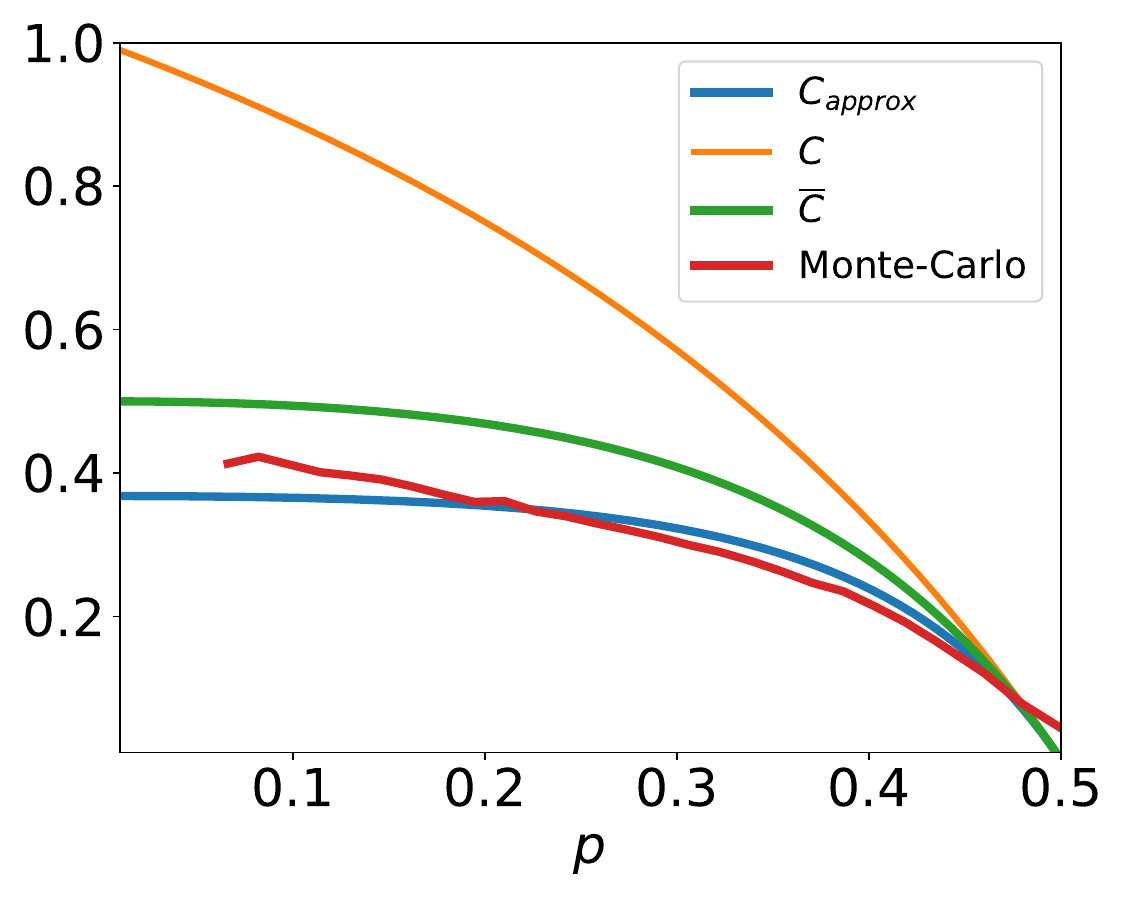}
\caption{The dependence of the cluster density  of the ultimate jammed state of a finite system on $p<1/2$. Bottom curve: The approximation \eqref{C-approx}. 
Middle curve: The approximation \eqref{C-bar}. For comparison, the cluster density $C=\frac{1-2p}{1-p}$ of the steady state of the infinite system is also shown 
(top curve). Our numerical results seem to be closer to $C_{\text{approx}}$ given by Eq.~\eqref{C-approx}. }
\label{Fig:CCC-approx}
\end{figure}

Another possible approximation relies on Eqs.~\eqref{123} uses the steady state \eqref{ckc:final} as the initial condition. Thus we postulate that suddenly all collisions lead to addition. Solving \eqref{3} and  \eqref{1} subject to the `initial condition' \eqref{CC1-0} gives
\begin{subequations}
\begin{align}
\label{c:SS}
&c(\tau) = \frac{1-2p}{1-p}\,e^{-\tau} \\
\label{c1:SS}
&c_1(\tau) =  \frac{1-2p}{1-p}\left[\frac{1-2p}{1-p} -\tau\right]e^{-\tau}
\end{align}
\end{subequations}
Substituting $\tau_*=\frac{1-2p}{1-p}$ where $c_1(\tau_*)$ given by \eqref{c1:SS} vanishes, into \eqref{c:SS} gives an approximation
\begin{equation}
\label{C-approx}
C_\text{approx} = e^\frac{2p-1}{1-p}\,\frac{1-2p}{1-p}
\end{equation}
for the final cluster density. All the approximations \eqref{C-bar} and \eqref{C-approx} for the cluster density, as well as the cluster density in the steady state, are decreasing functions of $p$ vanishing when $p=1/2$, see Fig.~\ref{Fig:CCC-approx}.

\section{AC processes with linear rates}
\label{ap: AC_linear}

Here we consider AC processes with rates \eqref{A-rate-lin}--\eqref{C-rate-lin}. The behavior greatly depends on which of the two processes, addition or chipping, prevails. 

\subsection{Infinite system}
\label{subsec: AC_linear}

The governing equations read 
\begin{subequations}
\begin{align}
&\frac{d c_1}{d\tau}= -c_1 - 1 + \lambda(2c_2-c_1+1)
\label{1AC:mass}\\
&\frac{d c_k}{d\tau}=(k-1)c_{k-1}-(1+\lambda)kc_k + \lambda (k+1)c_{k+1}
\label{2AC:mass}
\end{align}
\end{subequations}
Equations \eqref{2AC:mass} are valid for all $k\geq 2$. 

In the steady state regime, $\lambda>1$, the densities quickly reach the universal (independent on the initial condition) values
\begin{equation}
\label{Ck-SS}
C_k = \frac{1-\lambda^{-1}}{k\lambda^{k-1}}
\end{equation}
These densities are found from Eqs.~\eqref{1AC:mass}--\eqref{2AC:mass} by setting the left-hand sides to zero. The amplitude in \eqref{Ck-SS} is fixed by mass conservation: $\sum_{k\geq 1} k C_k=1$. 

The stationary cluster density 
\begin{equation}
\label{C-SS}
C = -(\lambda-1)\ln(1-\lambda^{-1}) 
\end{equation}
is a monotonically increasing function of $\lambda$.

In the critical regime, $\lambda = 1$, Eqs.~\eqref{1AC:mass}--\eqref{2AC:mass} become
\begin{equation}
\label{2-mass:crit}
\frac{d c_k}{d\tau}=(k-1)c_{k-1}-2kc_k + (k+1)c_{k+1}
\end{equation}
We can use \eqref{2-mass:crit} for all integer $k\geq 1$. These equations admit a neat solution
\begin{equation}
\label{sol-mass:crit}
c_k(\tau) = \frac{\tau^{k-1}}{(1+\tau)^{k+1}}
\end{equation}
in the case of the mono-disperse initial condition. The exact solution \eqref{sol-mass:crit} appears in various subjects ranging from birth-death processes to exchange processes \cite{economy-Slava,migration,BK-exchange}.

Re-writing \eqref{sol-mass:crit} in terms of the physical time we obtain
\begin{equation}
c_k(t) = \frac{1}{(1+3t)^{2/3}}\left[1-\frac{1}{(1+3t)^{1/3}}\right]^{k-1}
\end{equation}
In particular, the density of monomers and the cluster density are given by neat formulae
\begin{equation}
\label{c1c:mass}
c_1(t) = \frac{1}{(1+3t)^{2/3}}, \qquad c(t) = \frac{1}{(1+3t)^{1/3}}
\end{equation}

It appears impossible to obtain explicit results in the jamming regime, $\lambda<1$, so we limit ourselves with the most interesting asymptotic analysis just below the critical point: $0<1-\lambda\ll 1$. We  treat $\epsilon=1-\lambda$ as the small parameter and seek a perturbative solution 
\begin{equation}
c_k(\tau) = \frac{\tau^{k-1}}{(1+\tau)^{k+1}}+\epsilon f_k(\tau) + O(\epsilon^2)
\end{equation}
Plugging this expansion into \eqref{1AC:mass}--\eqref{2AC:mass} we obtain
\begin{subequations}
\begin{align}
&\frac{d f_1}{d\tau} = 2f_2 - 2f_1 -1-\frac{1}{(1+\tau)^2}+\frac{2}{(1+\tau)^3}
\label{f1:eq}\\
&\frac{d f_k}{d\tau} = (k+1)f_{k+1}-2kf_k + (k-1)f_{k-1} \nonumber \\
                             &\qquad\quad\frac{k\tau^{k-1}}{(1+\tau)^{k+1}}-\frac{(k+1)\tau^{k}}{(1+\tau)^{k+2}}
\label{fk:eq}
\end{align}
\end{subequations}
The analysis becomes feasible in the scaling regime
\begin{equation}
\label{scaling}
k\to\infty, ~~\tau\to\infty, ~~\xi=\frac{k}{\tau}=\text{finite}
\end{equation}
The infinite system \eqref{fk:eq} of ordinary differential equations turns into a single partial differential equation
\begin{equation}
\label{fk:PDE}
\frac{\partial f_k}{\partial \tau} = \frac{\partial^2}{\partial k^2}\,(kf_k) + \tau^{-2}(\xi-1)e^{-\xi}  
\end{equation}
in the scaling limit \eqref{scaling}. Seeking the solution of \eqref{fk:PDE} in the scaling form
\begin{equation}
\label{scaling:ansatz}
f_k(\tau) = \frac{1}{k}\,F(\xi)
\end{equation}
we recast \eqref{fk:PDE} into an ordinary differential equation
\begin{equation}
\label{Phi:ODE}
-\frac{d F}{d \xi} = \frac{d^2 F}{\partial \xi^2} + (\xi-1)e^{-\xi}  
\end{equation}
This equation admits a one-parameter family of solutions $F =  \left(\frac{\xi^2}{2}-C\right)e^{-\xi}$ vanishing when $\xi\to\infty$. Using \eqref{f1:eq} we fix the constant $C=1$. Thus
\begin{equation}
\label{Phi:sol}
F =  \left(\frac{\xi^2}{2}-1\right)e^{-\xi}  
\end{equation}
As a check of consistency let us compute the mass density. The correction term $\sum\limits_{k\geq 1}kf_k(\tau)\simeq \tau\int_0^\infty d\xi\,F(\xi)$, and using \eqref{Phi:sol} we find that it vanishes as it must. 

The cluster density  
\begin{equation}
c(\tau) = \frac{1}{1+\tau}+\epsilon\sum_{k\geq 1}f_k
\end{equation}
simplifies to
\begin{equation}
c(\tau) = \frac{1}{1+\tau}+\epsilon\left[\frac{1}{2}-E_1(1/\tau)\right]
\end{equation}
when $\tau\gg 1$. Here $E_1(z)=\int_z^\infty \frac{d\xi}{\xi}\,e^{-\xi}$ is the exponential integral. The monomer density is $c_1=(1+\tau)^{-2}-\epsilon$, from which $1+\tau_\text{max}=\epsilon^{-1/2}$. 
Therefore the final cluster density $C=c(\tau_\text{max})$ is 
\begin{equation}
C = \sqrt{\epsilon}+\frac{1}{2}\,\epsilon\ln(\epsilon)+O(\epsilon)
\end{equation}

\begin{figure}
\centering
\includegraphics[width=7.89cm]{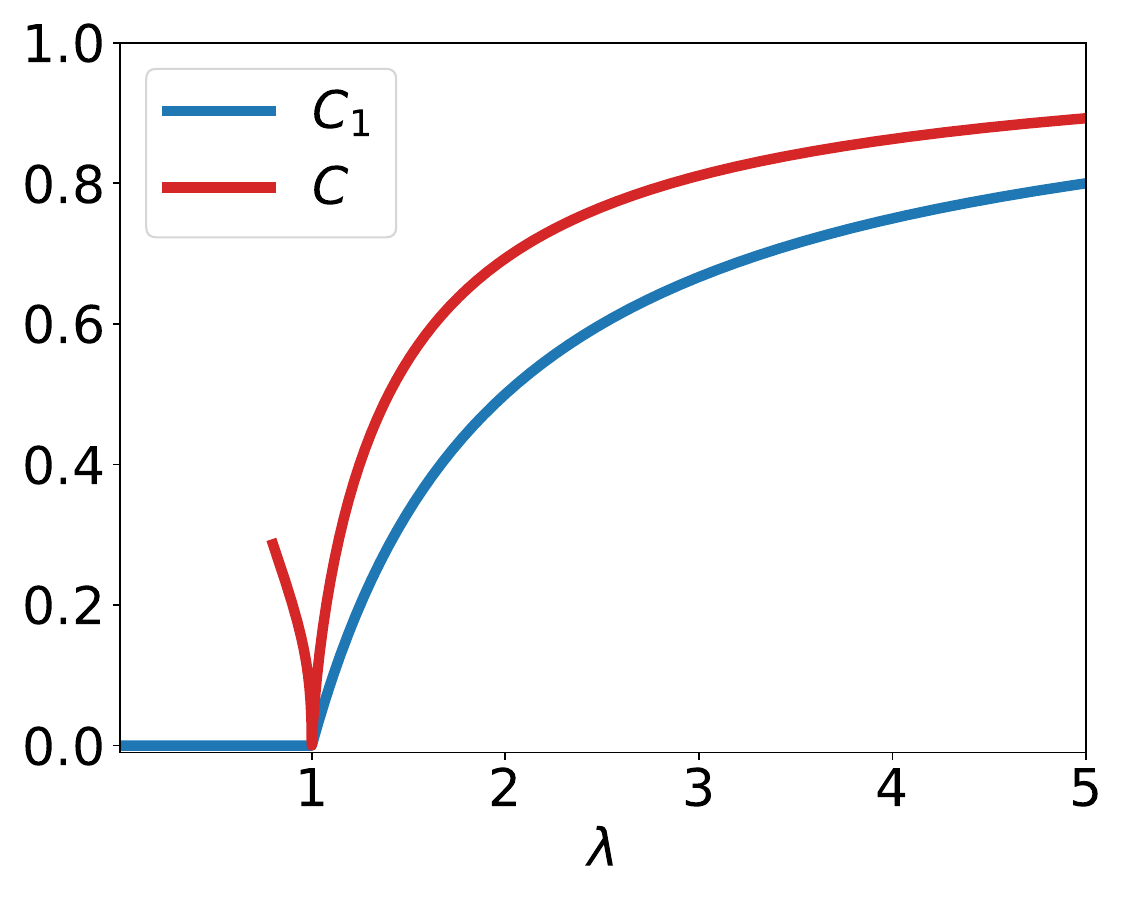}
\caption{Top curves: The final cluster density $C$. Bottom curves: The final monomer density $C_1$. The monomer density is known, Eq.~\eqref{mon-linear}, in the entire $0\leq \lambda<\infty$ range. The cluster density,  Eq.~\eqref{final:clusters}, is known in the steady state regime and in the jamming regime close to the critical point.  }
\label{Fig:CC1}
\end{figure}

Summarizing, in the entire range $0\leq \lambda<\infty$ of the chipping rate, the final density of monomers is given by  (see also Fig.~\ref{Fig:CC1})
\begin{equation}
\label{mon-linear}
C_1 = 
\begin{cases}
0                                        &\lambda \leq  1\\
1-\lambda^{-1}                  &\lambda > 1
\end{cases}
\end{equation}
This follows from \eqref{Ck-SS} at $k=1$ and illustrates the phase transition at $\lambda=\lambda_c=1$. 
The final total cluster density also undergoes a continuous phase transition: 
\begin{equation}
\label{final:clusters}
C = 
\begin{cases}
\sqrt{1-\lambda} + \frac{1-\lambda}{2} \ln(1-\lambda)+\ldots         &1-\lambda \ll 1\\
0                                                  &\lambda = 1\\
-(\lambda-1)\ln(1-\lambda^{-1})   &\lambda > 1
\end{cases}
\end{equation}

\subsection{Finite Systems}
\label{subsec:finite-linear}

A finite system gets jammed, but the scaling of the jamming time on $N$ greatly depends on whether $\lambda$ is smaller or larger than $\lambda=1$ when the addition and chipping processes balance. (Chipping is impossible when a monomer hits another monomer, and this imbalance drives evolution in the critical regime.) To appreciate the announced scaling laws \eqref{lifetime-linear}, we note that the steady state of an infinite system becomes quasi-stationary. Namely, the densities of a finite system fluctuate around the steady-state densities of an infinite system, but eventually, monomers disappear in a giant fluctuation. As in many other problems, the time for such highly improbably event scales exponentially (see \cite{Freidlin,VanKampen,KRB,Meerson17}), explaining the scaling law in \eqref{lifetime-linear} in the $\lambda>1$ region. The computation of the amplitude $D(\lambda)$ requires understanding giant fluctuations leading to the disappearance of monomers in a system with many interacting cluster species. Such computations are beyond the reach of available techniques \cite{Meerson17}).

The logarithmic scaling of the evolution time in the jamming regime, $\lambda < 1$, is an outcome of  an exponential relaxation in the infinite system. The critical regime is characterized by an algebraic evolution, so an algebraic dependence on the lifetime on $N$ is not surprising. The derivation of the scaling law announced in \eqref{lifetime-linear} is again subtle as the late stage is fluctuation-dominated. A naive argument relying on the decay law $c_1(T) \sim T^{-2/3}$ in the critical regime [cf. Eq.~\eqref{c1c:mass}] and the criterion $N  c_1(T)= O(1)$ gives $T\sim N^{3/2}$ which is erroneous. The correct answer,  $T\sim N^{3/4}$, relies on the analysis of fluctuations.  We will see (Sec.~\ref{sec:ApF}) that the stochastic part $\xi_1$ in \eqref{M:def} is of order one, $\langle \xi_1^2\rangle = O(1)$, in the late stage of the evolution. Monomers disappear when the deterministic and stochastic parts in \eqref{M:def} become comparable. Equating the deterministic part $Nc_1\sim N/T^{2/3}$ to the stochastic part $\sqrt{N}\sqrt{\langle \xi_1^2\rangle}\sim \sqrt{N}$ we obtain the scaling of the average lifetime
\begin{equation}
\label{time:crit}
T\sim N^\frac{3}{4}
\end{equation}
The final number of clusters is non-extensive
\begin{equation}
\label{ap:clusters:crit}
\mathcal{C} \sim  N c(T) \sim N T^{-1/3} \sim N^\frac{3}{4}  
\end{equation}
and the typical cluster mass is algebraically growing
\begin{equation}
\label{ap:mass:crit}
k_\text{typ} \sim T^\frac{1}{3} \sim N^\frac{1}{4}  
\end{equation}
These behaviors justify the name, the supercluster state, for the final state in the critical regime.

\subsection{Fluctuations in the critical regime}
\label{sec:ApF}

We use the same notations and the same procedure as in Sec.~\ref{sec:F}. The analog of \eqref{Exact} reads
\begin{subequations}
\label{ap:Exact}
\begin{align}
\label{ap:channel:m}
& (\mathcal{C}_1-2, \mathcal{C}_2+1)     ~\quad \qquad \qquad    \mathcal{C}_1(\mathcal{C}_1-1)/N \\
\label{ap:channel:add}
&(\mathcal{C}_1-1,\mathcal{C}_k-1,\mathcal{C}_{k+1}+1)              \quad k\mathcal{C}_1 \mathcal{C}_k/N  \\
\label{ap:channel:d}
& (\mathcal{C}_1+2, \mathcal{C}_2-1)     ~\quad \qquad \qquad     2\mathcal{C}_1\mathcal{C}_2/N    \\
\label{ap:channel:chip} 
&(\mathcal{C}_1+1,  \mathcal{C}_{k-1}+1  ,   \mathcal{C}_k-1)       \quad k\mathcal{C}_1 \mathcal{C}_k/N
\end{align}
\end{subequations}
The last reaction channel \eqref{ap:channel:chip} describes the chipping of clusters with $k\geq 3$; the chipping process with $k=2$ is represented by \eqref{ap:channel:d}.

Using Eqs.~\eqref{ap:Exact} we deduce equations for the averages
\begin{subequations}
\label{ap:Av-eqs}
\begin{align}
\label{ap:av-1-eq}
N\,\frac{d \langle \mathcal{C}_1\rangle}{dt} &= - 2 \langle \mathcal{C}_1(\mathcal{C}_1-1)\rangle +  2\langle \mathcal{C}_1 \mathcal{C}_2\rangle \\
\label{ap:av-2-eq}
N\,\frac{d \langle \mathcal{C}_2\rangle}{dt} &=  \langle \mathcal{C}_1(\mathcal{C}_1-1)\rangle -  4\langle \mathcal{C}_1 \mathcal{C}_2\rangle 
                                                                             +3\langle \mathcal{C}_1 \mathcal{C}_3\rangle 
\end{align}
and
\begin{eqnarray}
\label{ap:av-k-eq}
N\,\frac{d \langle \mathcal{C}_k\rangle}{dt} &=&  (k-1)\langle \mathcal{C}_1\mathcal{C}_{k-1}\rangle -  2k\langle \mathcal{C}_1 \mathcal{C}_k\rangle \nonumber \\
                                                                      &+& (k+1)\langle \mathcal{C}_1 \mathcal{C}_{k+1}\rangle 
\end{eqnarray}
\end{subequations}
for $k\geq 3$. 

Using the van Kampen expansions \eqref{M:def} and \eqref{Ck:def} together with identities \eqref{CCCk} we deduce
\begin{subequations}
\label{ap:xi-123}
\begin{align}
\label{ap:xi-1}
\frac{d \langle \xi_1\rangle}{dt}  & = -4 c_1 \langle \xi_1\rangle + 2[c_1\langle \xi_2\rangle+ c_2\langle \xi_1\rangle] \\
\label{xi-2}
\frac{d \langle \xi_2\rangle}{dt} &= 2 c_1 \langle \xi_1\rangle -4[c_1\langle \xi_2\rangle+ c_2\langle \xi_1\rangle] \nonumber\\
&+ 3[c_1\langle \xi_3\rangle+ c_3\langle \xi_1\rangle]
\end{align}
from \eqref{ap:av-1-eq} and \eqref{ap:av-2-eq}, while Eqs.~\eqref{ap:av-k-eq} give
\begin{eqnarray}
\label{ap:xi-3}
&&\frac{d \langle \xi_k\rangle}{dt} = c_1 [(k-1)\langle \xi_{k-1}\rangle -2k\langle \xi_k\rangle+ (k+1)\langle \xi_{k+1}\rangle] \nonumber \\
&&+\langle \xi_k\rangle [(k-1)c_{k-1}-2kc_k+(k+1)c_{k+1}]
\end{eqnarray}
\end{subequations}
for $k\geq 3$.  The initial state is deterministic: $\xi_k(0)=0$ for all $k\geq 1$. An infinite homogeneous system \eqref{ap:xi-123} of linear equations with initial condition $\langle \xi_k(0)\rangle = 0$ has a trivial vanishing solution: $\langle \xi_k\rangle\equiv 0$. Hence the first and  second-order cumulants are again given by \eqref{ij-av}.

We now notice that $\langle \mathcal{C}_1^2\rangle$ obeys
\begin{eqnarray}
\label{ap:MM-eq}
N\,\frac{d \langle \mathcal{C}_1^2\rangle}{dt} &=& \langle \mathcal{C}_1(\mathcal{C}_1-1)(-4\mathcal{C}_1+4)\rangle \nonumber \\
&+&  \sum_{k\geq 2} k\langle \mathcal{C}_1 \mathcal{C}_k(-2\mathcal{C}_1+1)\rangle  \nonumber \\
&+&  2\langle \mathcal{C}_1 \mathcal{C}_2 (4\mathcal{C}_1+4)\rangle  \nonumber \\
&+&  \sum_{k\geq 3} k\langle \mathcal{C}_1 \mathcal{C}_k(2\mathcal{C}_1+1)\rangle
\end{eqnarray}
Each term on the right-hand side (top to bottom) corresponds to the corresponding reaction channel in \eqref{ap:Exact}. 

Massaging the right-hand side of \eqref{ap:MM-eq} we obtain
\begin{eqnarray}
\label{ap:MM-short}
N\,\frac{d \langle \mathcal{C}_1^2\rangle}{dt} &=& -4\langle \mathcal{C}_1(\mathcal{C}_1-1)^2\rangle - 2 \langle \mathcal{C}_1^2 \rangle \nonumber \\
&+&  2N \langle \mathcal{C}_1\rangle + 2 \langle \mathcal{C}_1 \mathcal{C}_2 (2\mathcal{C}_1+3)\rangle
\end{eqnarray}
Combining \eqref{ap:av-1-eq} and \eqref{ap:MM-short} we obtain
\begin{eqnarray}
\label{ap:M-var}
N\,\frac{d \langle \mathcal{C}_1^2\rangle_c}{dt} &=& 
4[\langle \mathcal{C}_1^2\rangle \langle \mathcal{C}_1\rangle-\langle \mathcal{C}_1^3\rangle] + 6\langle \mathcal{C}_1  \mathcal{C}_2\rangle 
+ 2\langle \mathcal{C}_1^2 \rangle\nonumber \\
&+&4[\langle \mathcal{C}_1^2  \mathcal{C}_2\rangle-\langle \mathcal{C}_1\rangle \langle \mathcal{C}_1  \mathcal{C}_2\rangle] + 2N \langle \mathcal{C}_1\rangle  \nonumber \\
&+&4\langle \mathcal{C}_1^2 \rangle_c   - 4 \langle \mathcal{C}_1\rangle
\end{eqnarray}
Inserting \eqref{ij-av} and  \eqref{112-av}  into \eqref{ap:M-var}, keeping the leading $O(N^2)$ terms and using the modified time variable give
\begin{equation}
\label{ap:W11-eq}
\frac{d W_{11}}{d \tau} =4W_{12}-4\left(2-\frac{c_2}{c_1}\right)W_{11}+ 2c_1 + 6c_2+2
\end{equation}

To derive the evolution equation for $W_{12} = \langle \xi_1 \xi_2\rangle$, we first write
\begin{eqnarray}
\label{ap:C12-eq}
N\,\frac{d \langle \mathcal{C}_1 \mathcal{C}_2\rangle}{dt} &=& \langle \mathcal{C}_1^3\rangle - 6 \langle \mathcal{C}_1^2 \mathcal{C}_2\rangle
+  2\langle \mathcal{C}_1 \mathcal{C}_2^2\rangle +  3\langle \mathcal{C}_1^2 \mathcal{C}_3\rangle \nonumber \\
&+&  3\langle \mathcal{C}_1 \mathcal{C}_3\rangle  -3 \langle \mathcal{C}_1^2\rangle +  2 \langle \mathcal{C}_1\rangle
\end{eqnarray}
which we combine with Eqs.~\eqref{ap:av-1-eq}--\eqref{ap:av-2-eq} to find
\begin{eqnarray}
\label{ap:C12-eq-c}
N\,\frac{d \langle \mathcal{C}_1 \mathcal{C}_2\rangle_c}{dt} &=& 3\langle \mathcal{C}_1 (\mathcal{C}_3-\mathcal{C}_1) \rangle  + \langle \mathcal{C}_1\rangle^2
-2\langle \mathcal{C}_1\rangle \langle \mathcal{C}_2\rangle  \nonumber \\
&+&2 \langle \mathcal{C}_1\rangle+ [\langle \mathcal{C}_1^3\rangle - \langle \mathcal{C}_1^2\rangle \langle \mathcal{C}_1\rangle] \nonumber \\
&-& 4[\langle \mathcal{C}_1^2 \mathcal{C}_2\rangle-\langle \mathcal{C}_1 \mathcal{C}_2\rangle \langle \mathcal{C}_1\rangle] \nonumber \\
&-& 2[\langle \mathcal{C}_1^2 \mathcal{C}_2\rangle -  \langle \mathcal{C}_1^2\rangle \langle \mathcal{C}_2\rangle]  \nonumber \\
&+&  2[\langle \mathcal{C}_1 \mathcal{C}_2^2\rangle -  \langle \mathcal{C}_1 \mathcal{C}_2\rangle  \langle \mathcal{C}_2\rangle]  \nonumber \\
&+&  3[\langle \mathcal{C}_1^2 \mathcal{C}_3\rangle - \langle \mathcal{C}_1 \mathcal{C}_3\rangle  \langle \mathcal{C}_1\rangle]  
\end{eqnarray}
Using already known terms in brackets appearing in Eqs.~\eqref{112-av} and \eqref{123-av}, the leading behavior of two more terms in brackets
\begin{equation*}
\begin{split}
\langle \mathcal{C}_1^2 \mathcal{C}_2\rangle-\langle \mathcal{C}_1 \mathcal{C}_2\rangle \langle \mathcal{C}_1\rangle &= c_1 W_{12}+c_2 W_{11}\\
\langle \mathcal{C}_1^2 \mathcal{C}_2\rangle -  \langle \mathcal{C}_1^2\rangle \langle \mathcal{C}_2\rangle &= 2 c_1 W_{12}
\end{split}
\end{equation*}
and keeping the leading $O(N^2)$ terms we reduce \eqref{ap:C12-eq-c} to 
\begin{eqnarray}
\label{ap:W12-eq}
\frac{d W_{12}}{d \tau} &=&\left(2-\frac{2c_2}{c_1} +\frac{3c_3}{c_1}\right)W_{11}
-8W_{12}+2W_{22} + 3W_{13}\nonumber \\
& -& 2c_1 -2c_2+3c_3
\end{eqnarray}
Since $c_1\simeq c_2\simeq c_3$ and all decay to zero, Eqs.~\eqref{ap:W11-eq} and ~\eqref{ap:W12-eq} simplify to
\begin{subequations}
\begin{align}
\label{W11-simple}
\frac{d W_{11}}{d \tau} &=4W_{12}-4W_{11}+2\\
\label{W12-simple}
\frac{d W_{12}}{d \tau} &=3W_{11}-8W_{12}+2W_{22} + 3W_{13}
\end{align}
\end{subequations}
Generally, $W_{ij}$ with $(i,j)\ne (1,1)$ satisfy linear homogeneous equations similar to \eqref{W12-simple}. Therefore $W_{ij} = O(1)$ in the long time limit. We mostly need $W_{11} = O(1)$ leading to the scaling laws \eqref{time:crit}--\eqref{ap:mass:crit}.

\section{AC processes with algebraic rates}
\label{ap: AC_alg}

Here we consider a class of AC processes with algebraic rates \eqref{AC-prop}--\eqref{Ak-alg}. We focus on the most interesting critical regime ($\lambda=\lambda_c=1$). 

\subsection{Infinite critical system}
\label{subsec: AC_alg}

The infinite set of governing equations 
\begin{equation}
\label{ck:a-crit}
\frac{d c_k}{d \tau}  = (k-1)^a c_{k-1}- 2k^a c_k + (k+1)^a c_{k+1}
\end{equation}
is applicable for all $k\geq 1$ if we set $c_0\equiv 0$. 

The models with $a=0$ and $a=1$ are explicitly solvable as was shown above. The models with $a<2$ admit a scaling analysis  \cite{BK-exchange}. In the scaling regime
\begin{equation}
\label{scaling:a}
k\to\infty, ~~\tau\to\infty, ~~\xi=k \tau^{-\beta}=\text{finite}
\end{equation}
one seeks the self-similar solution 
\begin{equation}
\label{scaling:a-form}
c_k(\tau) = \tau^{-2\beta} \Phi(\xi), \quad \beta = (2-a)^{-1}
\end{equation}
Plugging \eqref{scaling:a-form} into \eqref{ck:a-crit} yields a differential equation for $\Phi(\xi)$ which is solved to yield \cite{BK-exchange}
\begin{equation}
\label{Phi:a-sol}
\Phi(\xi) = A\, \xi^{1-a}\,\exp\!\left[-\beta^2 \xi^{1/\beta}\right], \quad A = \frac{\beta^{2\beta}}{\Gamma(\beta)}
\end{equation}
The amplitude in \eqref{Phi:a-sol} was fixed using mass conservation: $\sum_{k\geq 1}kc_k=\int_0^\infty d\xi\,\xi\,\Phi(\xi) = 1$.

The monomer density and the cluster density decay as
\begin{equation}
\label{c1c:a}
c_1(\tau) = A\, \tau^{-1-\beta}, \qquad c(\tau) = \beta^{-1}A\,\tau^{-\beta}
\end{equation}
in the long time limit. In terms of the physical time
\begin{equation}
\begin{split}
\label{c1c:time-a}
& c_1 = \gamma_a\, t^{-(3-a)/(5-2a)}, \quad c = \nu_a\,t^{-1/(5-2a)} \\
&\gamma_a = \frac{A}{\left(A\,\frac{5-2a}{2-a}\right)^\frac{2-a}{5-2a}}\,, \quad 
\nu_a = \frac{A(2-a)}{\left(A\,\frac{5-2a}{2-a}\right)^\frac{1}{5-2a}}
\end{split}
\end{equation}
We also note that the typical cluster mass increases as
\begin{equation}
\label{mass:a}
k_\text{typ} \sim  t^\frac{1}{5-2a} 
\end{equation}

Above scaling analysis of the critical regime holds when $a<2$. The critical AC process with $a=2$ also admits an analytical treatment and exhibits an intriguing multi-scaling behavior \cite{BK-exchange}. An instantaneous and complete gelation happens in the critical AC processes with rates \eqref{AC-prop}--\eqref{Ak-alg} and $a>2$: $c_k(t)=0$ for all $k\geq 1$ and any $t>0$; see \cite{BK-exchange} for proof. On the physical ground the rates cannot grow faster than linearly, i.e., $a\leq 1$. Furthermore, the AC processes with algebraic rates and $a>1$ also exhibit an instantaneous gelation in the jamming regime ($\lambda>1$), see \cite{BK91,Laurencot99}. To avoid these pathological behaviors, we assume that $a\leq 1$. 

\subsection{Finite critical systems}
\label{subsec: finite-alg}

Here we consider a critical finite system. In a reaction event, the configuration $\{\mathcal{C}_1, \mathcal{C}_2, \ldots, \mathcal{C}_N\}$ transforms into one of the following configurations
\begin{subequations}
\label{ap:conf-a}
\begin{align}
\label{ap:mon-a}
& (\mathcal{C}_1-2, \mathcal{C}_2+1)     ~\quad \qquad \qquad    \mathcal{C}_1(\mathcal{C}_1-1)/N \\
\label{ap:add-a}
&(\mathcal{C}_1-1,\mathcal{C}_k-1,\mathcal{C}_{k+1}+1)              \quad k^a\mathcal{C}_1 \mathcal{C}_k/N  \\
\label{ap:d-a}
& (\mathcal{C}_1+2, \mathcal{C}_2-1)     ~\quad \qquad \qquad     2^a\mathcal{C}_1\mathcal{C}_2/N    \\
\label{ap:chip-a} 
&(\mathcal{C}_1+1,  \mathcal{C}_{k-1}+1  ,   \mathcal{C}_k-1)       \quad k^a\mathcal{C}_1 \mathcal{C}_k/N
\end{align}
\end{subequations}
The reaction channel \eqref{ap:chip-a} describes the chipping of clusters with $k\geq 3$; the chipping process with $k=2$ is represented by \eqref{ap:d-a}.

Repeating the same steps as in Sec.~\ref{sec:F} and Appendix~\ref{sec:ApF} we find the evolution equation for the average total number of monomers 
\begin{align}
\label{ap:av-1-a}
N\,\frac{d \langle \mathcal{C}_1\rangle}{dt} = - 2 \langle \mathcal{C}_1(\mathcal{C}_1-1)\rangle +  2\langle \mathcal{C}_1 \mathcal{C}_2\rangle
\end{align}
and for the variance
\begin{eqnarray}
\label{ap:M-var-a}
N\,\frac{d \langle \mathcal{C}_1^2\rangle_c}{dt} &=& 
4[\langle \mathcal{C}_1^2\rangle \langle \mathcal{C}_1\rangle-\langle \mathcal{C}_1^3\rangle] + 3\cdot 2^a\langle \mathcal{C}_1  \mathcal{C}_2\rangle 
+ 2\langle \mathcal{C}_1^2 \rangle\nonumber \\
&+&2^{1+a}[\langle \mathcal{C}_1^2  \mathcal{C}_2\rangle-\langle \mathcal{C}_1\rangle \langle \mathcal{C}_1  \mathcal{C}_2\rangle]  \nonumber \\
& +&  2\sum_{k\geq 1} k^a \langle \mathcal{C}_1  \mathcal{C}_k\rangle  + 4\langle \mathcal{C}_1^2 \rangle_c   - 4 \langle \mathcal{C}_1\rangle
\end{eqnarray}
Keeping the leading $O(N^2)$ terms in \eqref{ap:M-var-a} and using the modified time variable we deduce
\begin{eqnarray}
\label{ap:W11-a}
\frac{d W_{11}}{d \tau} &=& 2^{1+a}W_{12}-2\left(4-\frac{2^a c_2}{c_1}\right)W_{11} \nonumber \\
& +& 2c_1 + 3\cdot 2^a c_2+2m_a
\end{eqnarray}
with $m_a= \sum_{k\geq 1}k^a c_k$. In the long time limit
\begin{eqnarray}
m_a  &\simeq&  \tau^{-\beta(1-a)} \int_0^\infty d\xi\,\xi^a \Phi(\xi) \nonumber \\
          & = &         \tau^{-\beta(1-a)}\,\frac{\beta^{1-2\beta} \Gamma(2\beta)}{\Gamma(\beta)}  \sim  t^{-\frac{1-a}{5-2a}}
\end{eqnarray}
In the long time limit $c_1, c_2\ll m_a$ and $c_2/c_1\to 2^{1-a}$, so \eqref{ap:W11-a} simplifies to
\begin{equation}
\label{W11-simple-a}
\frac{d W_{11}}{d \tau} =2^{1+a}W_{12}-4W_{11}+2m_a
\end{equation}

The same argument as before suggests $W_{11}\sim m_a$. The criterion\eqref{crit:NW} then gives the announced estimate \eqref{T-a} for $T$. Using \eqref{T-a} and \eqref{c1c:time-a}--\eqref{mass:a} we deduce the scaling of the final number of clusters and the typical cluster mass in the supercluster state
\begin{equation}
\label{number-mass:crit-a}
\mathcal{C} \sim N T^{-\frac{1}{5-2a}}  \sim N^\frac{4-a}{5-a}, \quad k_\text{typ} \sim  N^\frac{1}{5-a} 
\end{equation}

\bibliography{references-AC}

\end{document}